\documentclass[fleqn,usenatbib,useAMS]{mnras}
\usepackage{newtxmath}
\usepackage[T1]{fontenc}
\usepackage{ae,aecompl}

\usepackage{graphicx}   
\usepackage{amsmath}    
\usepackage{amssymb}    
\usepackage{pdflscape}  

\title[Supernova Correlations]{Temporal and Angular Variations of 3D Core-Collapse Supernova Emissions and their Physical Correlations}

\author[D. Vartanyan et. al]{David Vartanyan$^{1}$\thanks{E-mail: dvartany@princeton.edu},
Adam Burrows$^{1}$,
David Radice$^{1,2}$ \\
$^{1}$Department of Astrophysical Sciences, Princeton University, Princeton, NJ 08544\\
$^{2}$ Institute for Advanced Study, 1 Einstein Dr, Princeton NJ 08540\\
}

\begin{document}
\label{firstpage}
\maketitle

\begin{abstract}
We provide the time series and angular distributions of the neutrino and gravitational-wave emissions of eleven state-of-the-art three-dimensional non-rotating core-collapse supernova models and explore correlations between these signatures and the real-time dynamics of the shock and the proto-neutron-star core.  The neutrino emissions are roughly isotropic on average, with instantaneous excursions about the mean inferred luminosity of as much as $\pm$20\%. The deviation from isotropy is least for the ``$\nu_{\mu}$"-type neutrinos and the lowest-mass progenitors. Instantaneous temporal luminosity variations along a given direction for exploding models average $\sim$2$-$4\%, but can be as high as $\sim$10\%. For non-exploding models, they can achieve $\sim$25\%.  The temporal variations in the neutrino emissions correlate with the temporal and angular variations in the mass accretion rate. We witness the LESA phenomenon in all our models and find that the vector direction of the LESA dipole and that of the inner Y$_\mathrm{e}$ distribution are highly correlated. For our entire set of 3D models, we find strong connections between the cumulative neutrino energy losses, the radius of the proto-neutron star, and the $f$-mode frequency of the gravitational wave emissions. When physically normalized, the progenitor-to-progenitor variation in any of these quantities is no more than $\sim$10\%. Moreover, the reduced $f$-mode frequency is independent of time after bounce to better than $\sim$10\%. Therefore, simultaneous measurement of gravitational waves and neutrinos from a given supernova event can be used synergistically to extract real physical quantities of the supernova core.
\end{abstract}

\section{Introduction}
\label{sec:int}

Recent supernova code developments (e.g., \citealt{skinner2019}) running on a new generation of supercomputers have ushered in an era of unparalleled 3D simulations of core-collapse supernovae (CCSNe). For instance, our group is now able to produce over ten three-dimensional supernovae simulations in under a year, rivaling even the most comprehensive efforts in two-dimensions just a few years ago. With such a sample size, one can begin to perform preliminary statistics of explosion characteristics and outcomes.

The advent of such computational capability has been paralleled by novel astronomical detector capabilities. Recent discoveries of gravitational waves from astrophysical sources (\citealt{abott2016}) and existing and upcoming neutrino detectors (SuperKamiokande (\citealt{abe2016}), HyperKamiokande (\citealt{abe2011,abe2018}), DUNE (\citealt{migenda2018,ankowski2016}), JUNO (\citealt{lu2016}), IceCube (\citealt{abbasi2011,kopke2011}) motivate new efforts to scrutizine direct signatures of proto-neutron star formation and supernova explosion. Neutrino detection of a galactic supernova will provide insight into both the dynamics of explosion and the physics of matter at nuclear densities (\citealt{muller2019}). In addition, \cite{suwa2019} emphasize that, depending upon the neutron star mass, galactic supernova neutrinos can be observed for 30$-$100 seconds. Simultaneous observation of gravitational waves (GWs) will constrain PNS convection and $g$/$f$-mode oscillation (\citealt{vsg2018,hayama2018,radice2019}) as well as neutrino-driven convection and possibly the standing accretion shock instability (SASI) (\citealt{tamborra2013,tamborra2014,kuroda2017,walk2019}). CCSN gravitational waves are detectable for galactic events via the LIGO/Virgo network, and even further with future third-generation detectors (\citealt{srivasta2019,powell2019}).

Many groups are now capable of high-fidelity 3D simulations (\citealt{vartanyan2018b,radice2019,burrows_2019,hiroki_2019,oconnor_couch2018b,muller2017,summa2018,glas2019,takiwaki2016,takahashi2019,nagakura2019,roberts2016,ott2018_rel}).
Our code F{\sc{ornax}} is unique in its inclusion of detailed microphysics (including inelastic scattering), fast explicit 
transport with an implicit local solver (without the ray-by-ray approximation and with velocity dependence),  respectable 
angular and radial resolution (\citealt{hiroki_2019}), and static mesh refinement to obviate severe Courant limitations in the core and on the axis.  The result is a code sporting the necessary realism that is five$-$ten times faster than previous 3D codes. 

Using results from F{\sc{ornax}}, we present in this paper the first study of variability and correlation to include a sample of almost a dozen state-of-the-art core-collapse supernovae simulations done with sophisticated neutrino physics and transport. The objective of this paper is to correlate CCSNe observables with physical quantities in the core to provide a basis for using neutrino and gravitational measurements to constrain the physical phenomena in the opaque CCSNe core. Earlier such work on correlations (\citealt{totani1998,Raffelt_2005,brandt2011,tamborra2013,nakamura2016,wallace2016,kuroda2016,kuroda2017, seadrow2018,hayama2018,walk2019}) focused on far fewer models with either a sub-optimal suite of included physics or at lower dimensionality. We study the neutrino and gravitational wave signatures and temporal and spatial variations, as well as correlations with the CCSN progenitor, for a comprehensive suite of 11 progenitors spanning 9$-$60 M$_{\odot}$ evolved in 3D. These models, or a subset, were explored in three earlier works: \cite{vartanyan2018b}, where we investigated the explosion of a 16-M$_{\odot}$ in 3D;  \cite{burrows_2019}, where we studied the evolution of the 9-13-M$_{\odot}$ progenitors from this suite; in \cite{radice2019}, where we analyzed the gravitational wave signal of progenitors in this suite, and in \cite{hiroki_2019}, where we looked at the dependence of angular resolution on explosion outcome for the 19-M$_{\odot}$ progenitor. We save for an upcoming paper a broader study of the explosion characteristics and phenomenology of this large 3D model suite (Burrows et.\,al, in preparation). The so-called diagnostic explosion energies for most of our models have yet to asymptote to their final values.  The exception is the 9-M$_{\odot}$ model, which early achieved a value just above 10$^{50}$ ergs ($\sim$0.1 Bethe).  The diagnostic energies of all the other models at run termination are larger (in the range of $\sim$10$^{50}$ to a few$\times$ 10$^{50}$ ergs) and still increasing, with that for the 25-M$_{\odot}$ model increasing at the end of its run at a rate of $\sim$1 Bethe per second.

We organize the paper as follows: In \S\ref{sec:outline}, we outline the physical and numerical setup of our simulation. In \S\ref{sec:neutrino}, we explore the temporal and spatial variations of neutrinos. We investigate the time series of neutrino emissions accessible to future neutrino detectors, and identify the associated physical processes driving the time variability. In \S\ref{sec:GW}, we repeat the process for gravitational waves. In \S\ref{sec:SASI}, we comment on where and when the SASI might emerge. In \S\ref{sec:LESA}, we explore the possible presence of the LESA \citep{2014ApJ...792...96T} in our 3D simulations. In \S\ref{sec:corr}, we look at correlations between observable gravitational and neutrino signals and the physics of CCSNe, such as neutron star convection, shock radius growth, and accretion rate. In \S\ref{sec:conc}, we conclude with summary comments and conclusions.

\section{Numerical Setup}\label{sec:outline}

F{\sc{ornax}} is a multi-dimensional, multi-group radiation hydrodynamics code originally
constructed to study core-collapse supernovae. Its design, capabilities, and a variety of core tests are detailed in \cite{skinner2019}.
In 2D and 3D, F{\sc{ornax}} employs a dendritic grid which deresolves in angle at small radii to avoid restrictive CFL timestep limitations, while at the same time preserving cell size and 
aspect ratios.  Our method of deresolving near the polar axis for 3D simulations allows us partially to
overcome axial artifacts seen conventionally in 3D simulations in spherical coordinates 
(\citealt{2015ApJ...807L..31L,muller2017}). F{\sc{ornax}} solves the 
comoving-frame velocity-dependent transport equations to order O($v/c$). The hydrodynamics 
uses a directionally-unsplit Godunov-type finite-volume scheme and computes fluxes at cell 
interfaces using an HLLC Riemann solver. For all the 3D simulation highlighted in this paper, we employ a spherical grid in r, $\theta$, and $\phi$
of resolution 678$\times$128$\times$256 (608$\times$128$\times$256 for the 16-M$_{\odot}$ model).  
The radial grid extends out to 20,000 kilometers (10,000 for the 16-M$_{\odot}$ model) and is spaced evenly with $\Delta{r}\sim0.5$ km 
for radii interior to 20 km and logarithmically for radii exterior to 50 km, with a smooth transition in between. 
The angular grid resolution varies smoothly from $\Delta\theta \sim$1.9$^\circ$ at the poles 
to $\Delta\theta \sim$1.3$^\circ$ at the equator, and has $\Delta\phi \sim$1.4$^\circ$ uniformly. 
For this project, following \cite{marek2006} we used a monopole approximation for relativistic gravity
and employed the SFHo equation of state (\citealt{2013ApJ...774...17S}), which is 
consistent with all currently known nuclear constraints (\citealt{2017ApJ...848..105T}). 

We solve for radiation transfer using the M1 closure scheme for the second and 
third moments of the radiation fields (\citealt{2011JQSRT.112.1323V}) and follow 
three species of neutrinos: electron-type ($\nu_{e}$), anti-electron-type ($\bar{\nu}_{e}$), 
and ``$\nu_{\mu}$"-type ($\nu_{\mu}$, $\bar{\nu}_{\mu}$, $\nu_{\tau}$, and $\bar{\nu}_{\tau}$ 
neutrino species collectively). We use 12 energy groups spaced logarithmically between 1 and 300 MeV 
for the electron neutrinos and to 100 MeV for the anti-electron- and ``$\nu_{\mu}$"-neutrinos.  Our choice of energy binning for the different species reflects that, at the high densities in the inner core, electron neutrino transport dominates energy transport and the Fermi level of degenerate electron neutrinos
is high enough to require the energy grouping extend to high enough
values for them (in our calculations 300 MeV). Electron neutrino degeneracy in the core suppresses electron anti-neutrino densities there severely and so their energy grouping need extend to only 100 MeV, allow a more highly-resolved energy grid. Since the heavy-neutrinos are not degenerate and are subdominant energy carriers in the core, we employed the same narrower and more highly-resolved energy range.

F{\sc{ornax}} includes a detailed suite of microphysical reactions. For neutrino-nucleon scattering (neutral-current) and absorption (charged current), we use the formalism of \cite{2006NuPhA.777..356B}. For inelasticity in neutrino-nucleon and neutrino-electron scattering, we follow \cite{burrows_thompson2004}. Weak magnetism and recoil corrections are included in both scattering absorption reactions as multiplicative terms. Pauli exclusion is included via the stimulated emission term in \citep{2006NuPhA.777..356B}. We include the many-body correction (see also \citealt{burrows2018}) to neutrino-nucleon scattering following \cite{PhysRevC.95.025801}, which fits to high densities following \cite{1998PhRvC..58..554B}. For electron-positron pair annihilation and bremsstrahlung, which dominate heavy neutrino production, we follow \cite{thomp_bur_horvath}. For electron capture on nuclei, significant during the infall phase, we use the rates provided by \cite{2010NuPhA.848..454J}. The cross sections for the flavor-changing reactions (e.g., $\nu_e$ + $\bar{\nu}_e$ $\rightarrow$ $\nu_\mu$ + $\bar{\nu}_\mu$; see, for instance \citealt{buras2003}) are small compared to neutrino/nucleon absorption and scattering processes and are not included in the current implementation of F{\sc{ornax}}. However, (\citealt{kotake2018}) have recently suggested that they might play a role in enhancing the contraction of the PNS and, thereby, slightly increasing the important neutrino heating rate in the gain region.

We study 11 progenitors in 3D in this paper, covering 9-, 10-, 11-, 12-, 13-, 14-, 15-, 16-, 19-, 25-, and 60-M$_{\odot}$ models. All models in this paper are initially collapsed in 1D through 10 ms after bounce, and 
then mapped to three dimensions. For all progenitors except the 16- and 25-M$_{\odot}$ models, we use \cite{swbj16}. We use \cite{wh07} for the 16-M$_{\odot}$ progenitor, the same studied in \cite{vartanyan2018b}. For the 25-M$_{\odot}$ progenitor, we use \cite{sukhbold2018}. After mapping to 3D, we impose velocity perturbations following \cite{muller_janka_pert}
within 200 $-$ 1000 km with a maximum speed of 100 km s$^{-1}$ and harmonic quantum numbers of $l$ = 10, $m$ = 1, and $n$ = 4 (radial), as defined in \cite{muller_janka_pert}, for all models except the 16-M$_{\odot}$ progenitor, which is perturbed in three spatially distinct regions (50 $-$ 85 km, 90 $-$ 250 km, and 260 $-$ 500 km),  
with a maximum speed of 500 km s$^{-1}$ and harmonic quantum numbers of $l$ = 2, $m$ = 1, 
and $n$ = 5 (radial). The details of the imposed perturbations are unlikely to make any qualitative difference in our 
conclusions.

\begin{figure}
    \centering
    \includegraphics[width=0.5\textwidth]{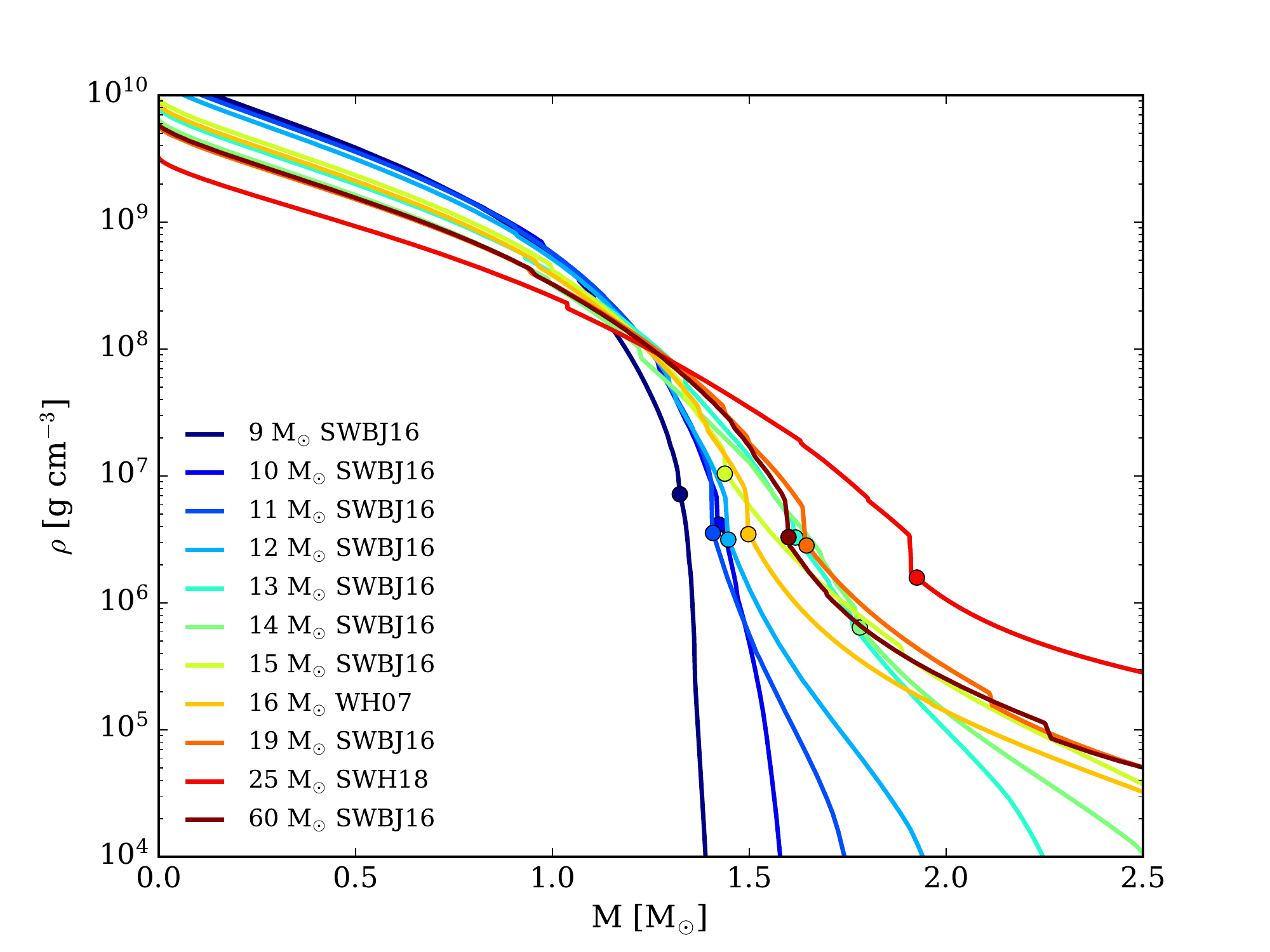}
    \caption{Mass density profile (in g cm$^{-3}$) with interior mass (in M$_{\odot}$) for the 11 progenitors of this study from three different model sets, studied in this series. The labels SWBJ16, WH07, and SWH18 stand for the progenitor models of \protect\cite{swbj16}, \protect\cite{wh07}, and \protect\cite{sukhbold2018}, respectively. The location of the sharp density drop at the silicon-oxygen interface, whose accretion often inaugurates explosion, is marked as a circle for each progenitor.}
    \label{fig:rho_M}
\end{figure}

In Fig.\,\ref{fig:rho_M}, we plot the density profiles as a function of mass for the progenitors studied here.  We highlight the accretion of the silicon-oxygen (Si/O) interface, often corresponding to a drop by several in density, with a colored diamond. Density profiles historically have been parametrized by compactness (\citealt{oconnor2013}), defined at a given mass (typically 1.75 M$_{\odot}$) and radius. However, we note the wide diversity of locations for the Si/O interfaces in this progenitor suite, from interior masses of 1.3 to 1.9 M$_{\odot}$.  Accretion of this interface often coincides with explosion time, and its presence and significance would require 
a multi-dimensional parametrization of progenitor profiles.

\begin{figure}
    \centering
    \includegraphics[width=0.5\textwidth]{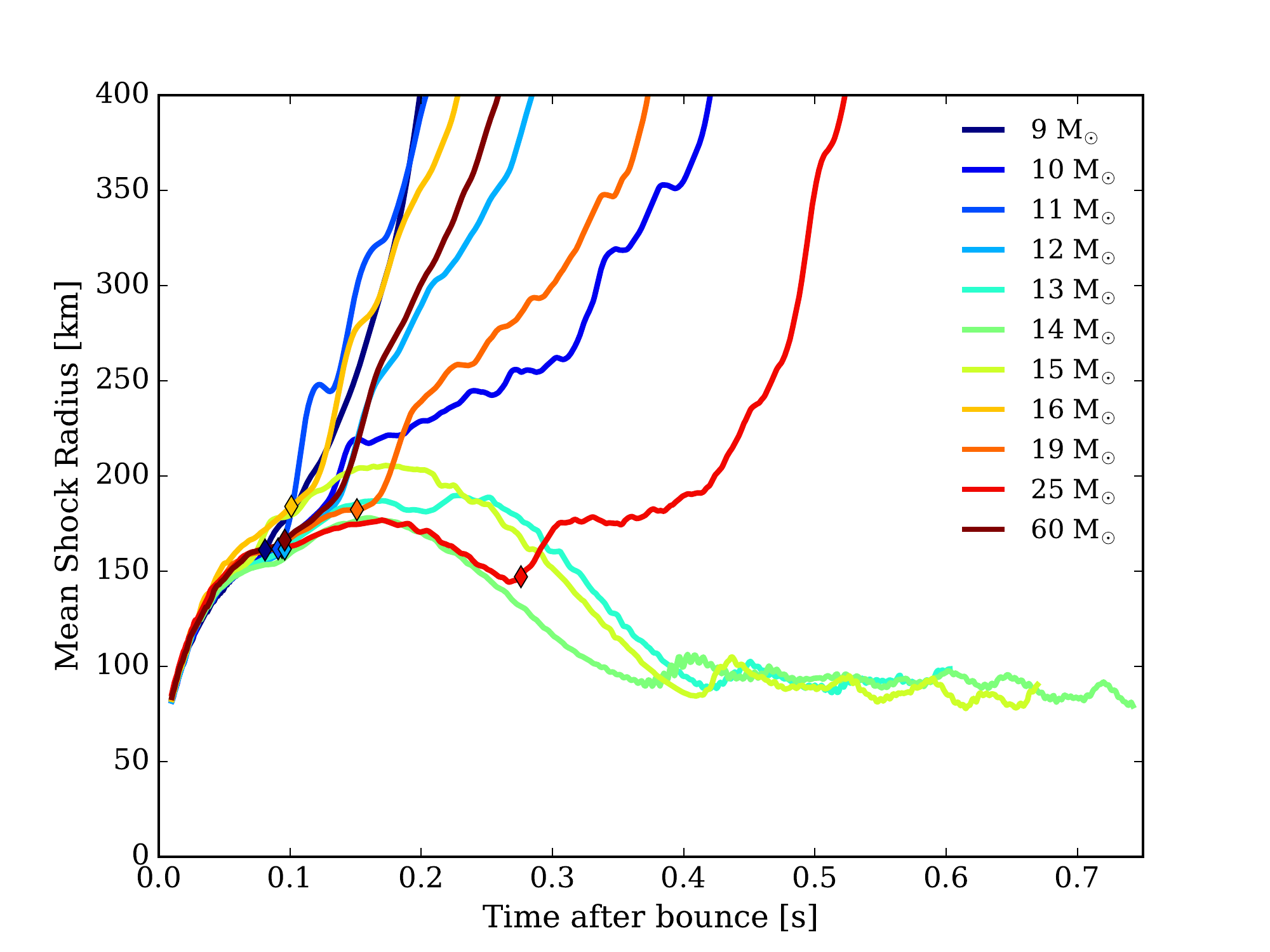}
    \caption{Mean shock radius (in km) as a function of time after bounce (in seconds) for the progenitors studied in this paper. The  diamonds indicate the approximate onset of explosion; all models except the 13-, 14-, and 15-M$_{\odot}$ progenitors explode. The 25-M$_{\odot}$ progenitor explodes latest, at $\sim$275 ms postbounce. Explosion time here corresponds closely with the accretion of the Si/O interface.}
    \label{fig:rshock}
\end{figure}

In Fig.\,\ref{fig:rshock}, we illustrate the evolution of the mean shock radii for all our 3D models. All progenitors except for the 13-, 14-, and 15- M$_{\odot}$ models explode, where we see an island of non-explosion (\citealt{burrows_2019}) and a weaker Si/O interface. Previously, the 15-M$_{\odot}$ has either failed to explode in 2D simulations (\citealt{vartanyan2018a, oconnor_couch2018a}) or has exploded late (\citealt{summa2016}). We emphasize that the 13-, 14-, and 15-M$_{\odot}$ progenitors are definitely less explodable. However, those of our models that do not explode may do so with different physics or initial structure, such as an initially rotating progenitor, or at higher resolution (\citealt{hiroki_2019}).

For specificity, we define explosion time as when the mean shock radius surpasses 150 km and undergoes an inflection point. Explosion sets in primarily between 100$-$200 ms after bounce except for the the 25-M$_{\odot}$ progenitor, which explodes near 275 ms after bounce, due to delayed accretion of the Si/O interface that is initially located further out. This model is unique in that we see shock revival at $\sim$275 ms, then a period of slow growth until $\sim$400 ms, after which the mean shock radius accelerates outwards. The explosion times we witness roughly correspond with the times of accretion of the Si/O interfaces (seen in Fig.\,\ref{fig:rho_M}), with a deeper steep Si/O interface predicting an earlier explosion time (\citealt{fryer1999,murphy2008}). More recent simulations have identified the importance of accretion of the Si/O interface in prompting earlier explosion in both 2D (\citealt{radice2017b,suwa2016,vartanyan2018a}) and 3D simulations (\citealt{hanke2013,ott2018_rel,summa2018,vartanyan2018b,burrows_2019}).

\section{Temporal and Directional Variation of Neutrino Emission}\label{sec:neutrino}

\begin{figure}
    \centering
    \includegraphics[width=0.5\textwidth]{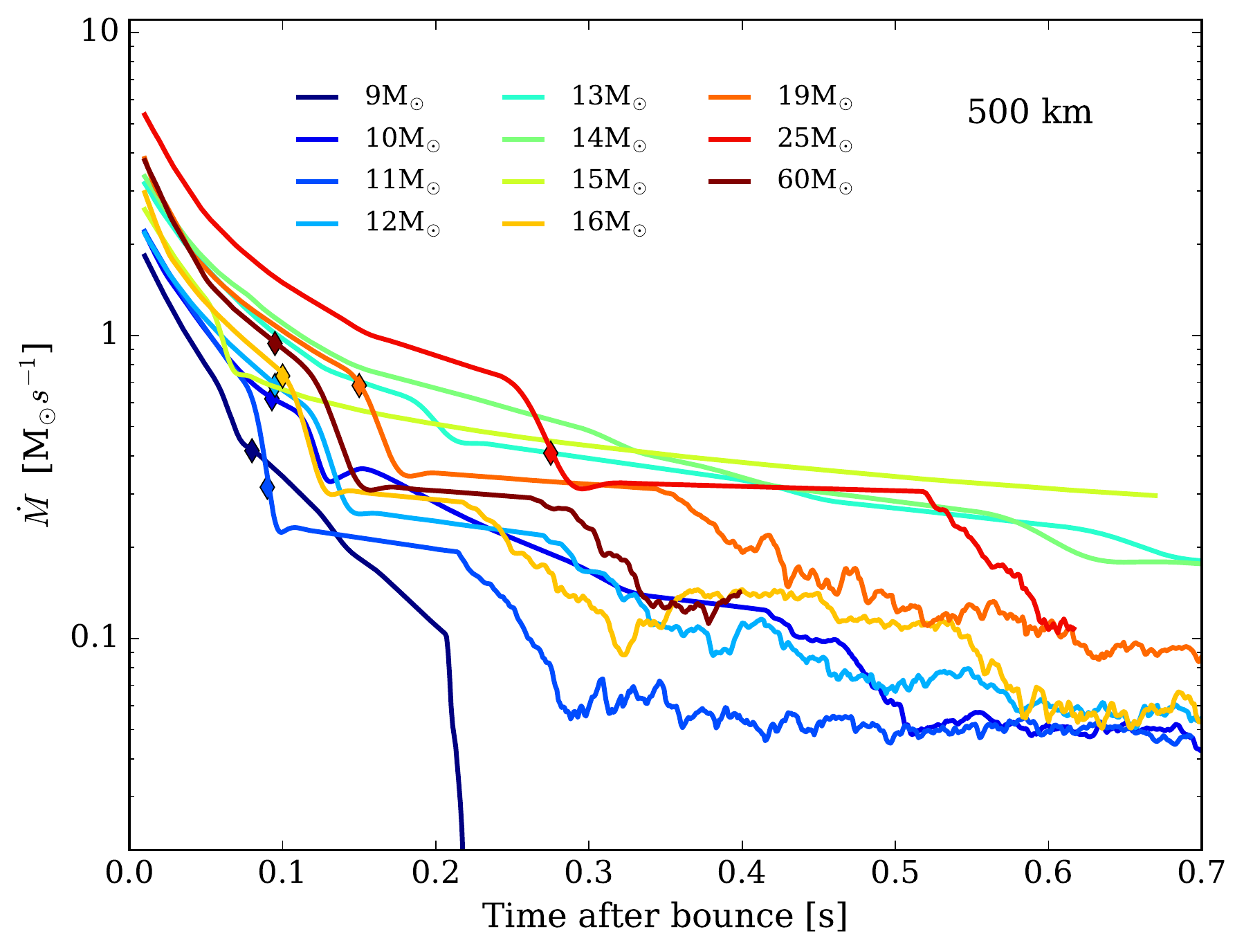}
    \caption{The accretion rate (M$_{\odot}$ s$^{-1}$) at 500 km as a function of time after bounce (in seconds). The accretion rate increases in general with core compactness. The drop in the accretion rate corresponds to the sharp density drop at the infalling Si/O layer. The times of explosion are marked as diamonds. Note the steep drop in accretion rate for the 9-M$_{\odot}$ progenitor just after 200 ms.}
    \label{fig:mdot}
\end{figure}

\begin{figure}
    \centering
    \includegraphics[width=0.5\textwidth]{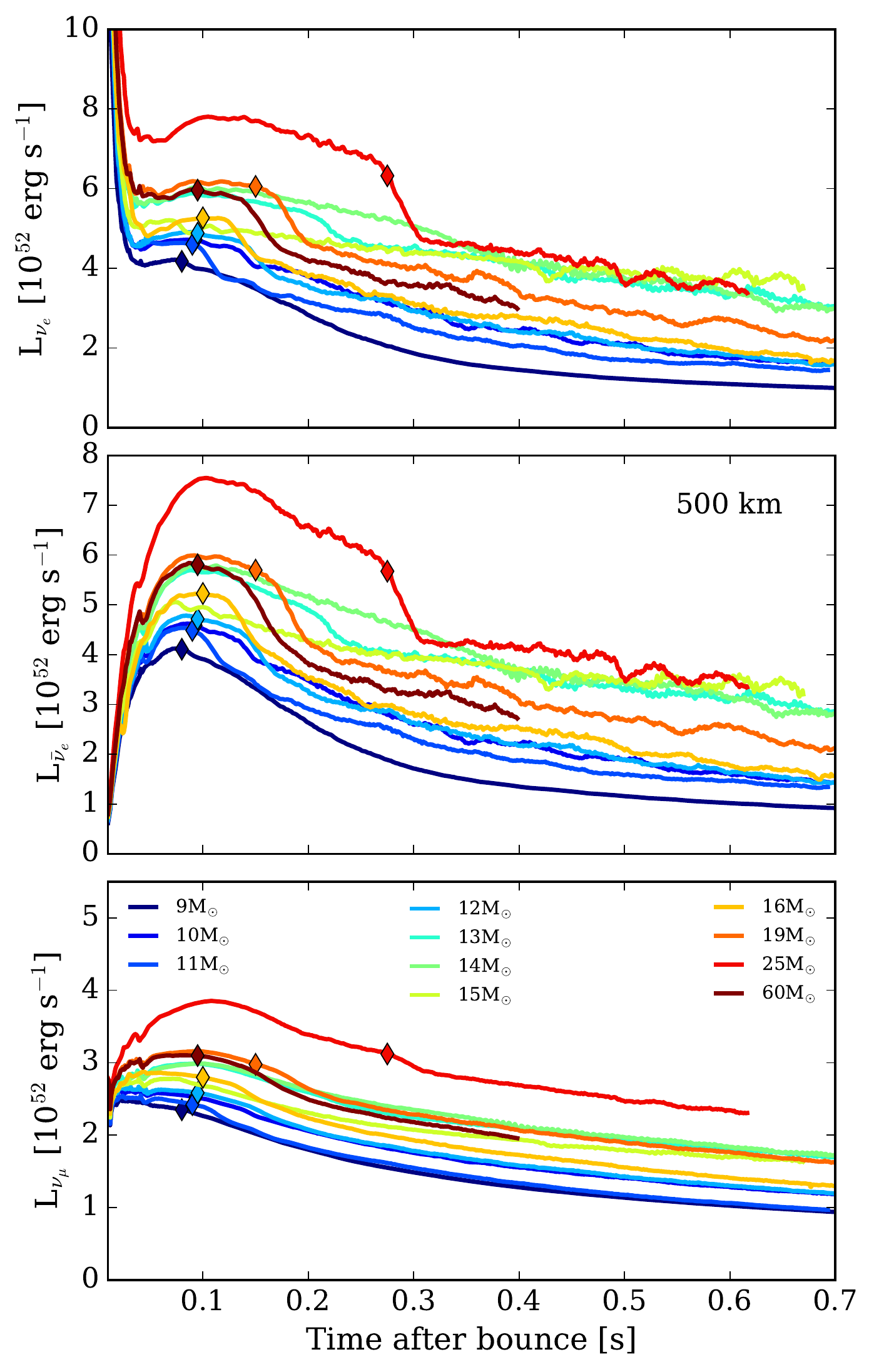}
    \caption{Mean electron neutrino (top), electron anti-neutrino  (middle), and the bundled heavy-neutrino (bottom) luminosities (in 10$^{52}$ erg s$^{-1}$) as a  function of time after bounce (in seconds). The accretion of the Si/O interface by the shock and ensuing explosion (diamonds) correspond to a sharp drop in the electron neutrino and anti-neutrino luminosities, with a more subdued drop in the heavy-neutrino luminosities. We note that the luminosity increases roughly with increasing progenitor mass (the 60-M$_{\odot}$ is an outlier) and closely with core compactness (see Fig.\,\ref{fig:compact}.)}
    \label{fig:lum}
\end{figure}

\begin{figure}
    \centering
    \includegraphics[width=0.5\textwidth]{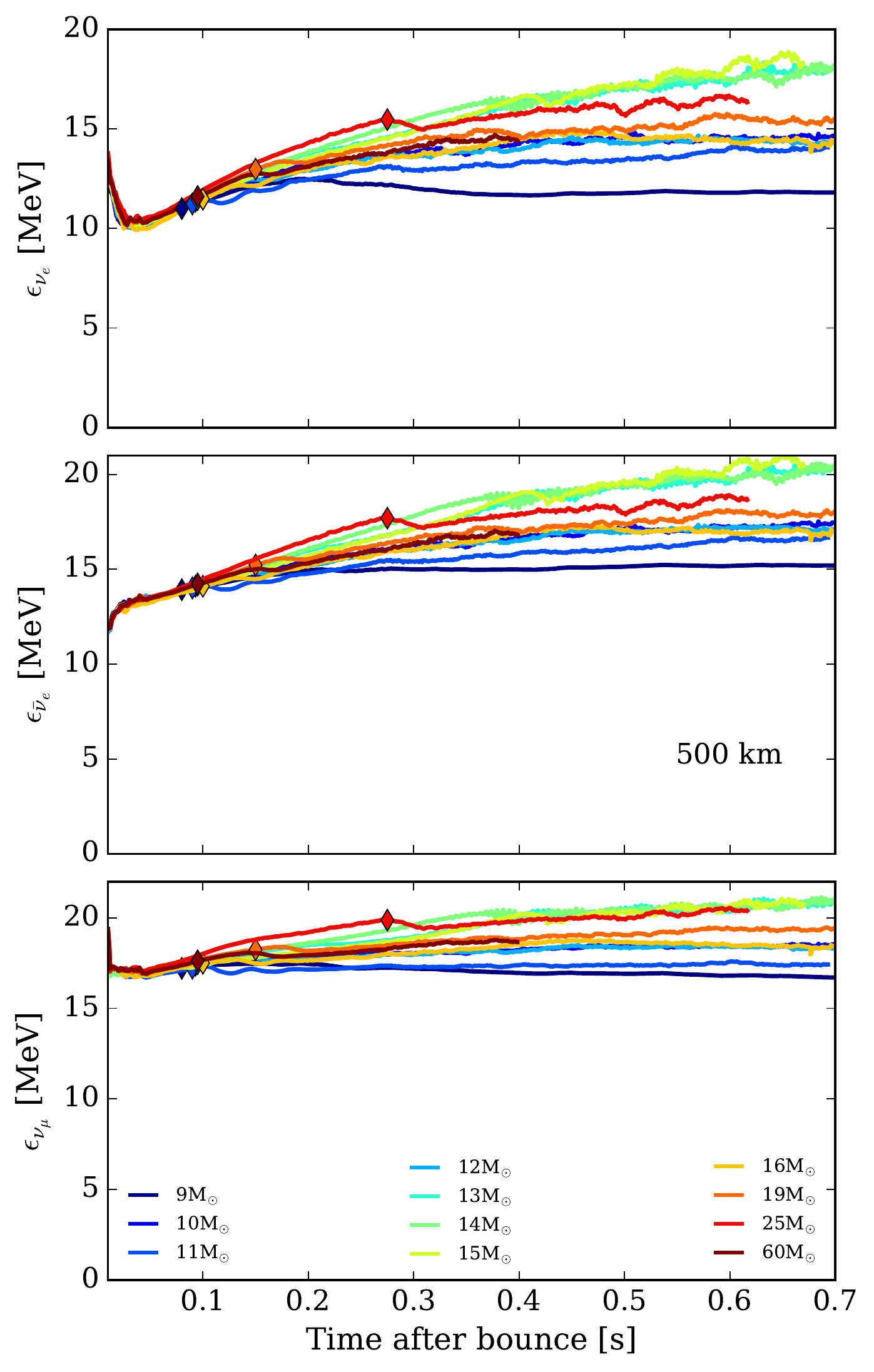}
    \caption{Electron neutrino (top), electron anti-neutrino  (middle), and the bundled heavy-neutrino (bottom) mean energies in MeV as a function of time after bounce (in seconds). The explosion time is marked in diamonds. The early turnover in the exploding models in neutrino energy corresponds to the accretion of the Si/O interface. Note that the non-exploding models have higher mean energies after $\sim$300 ms as a result of sustained accretion.  The mean energies flatten out at late times.}
    \label{fig:energy}
\end{figure}

We now explore the neutrino emission in time and angle. In Fig.\,\ref{fig:mdot}, we plot the accretion rates at 500 km as a function of time after bounce out to 700 ms for our suite of 3D models. In Fig.\,\ref{fig:lum} and \ref{fig:energy}, we plot the solid-angle-averaged neutrino luminosities and mean energies, respectively, at 500 km as a function of time after bounce for all eleven models. The $\nu_\mu, \nu_\tau$ neutrinos and anti-neutrinos are bundled into ``heavy"-neutrinos in the plots. The heavy neutrinos have slightly higher mean energies ($\sim$5\%) than the electron anti- neutrinos, which in turn have slightly higher neutrino energies ($\sim$15\%) than the electron-neutrinos. The average neutrino luminosities are not in equipartition between species, nor do they exhibit a strict hierarchy by species. Furthermore, the summed electron-neutrino and anti-electron-neutrino luminosity is roughly equal to the total heavy-neutrino luminosity. (\citealt{Raffelt_2005,totani1998}).

The models that fail to explode have a longer sustained accretion history and yield at late times, after $\sim$300 ms, higher accretion rates, mean neutrino energies, and neutrino luminosities, with observable consequences for neutrino detectors (\citealt{seadrow2018}).  The neutrino luminosities and mean energies flatten out at late times with the cessation of accretion. We note that the 9-M$_{\odot}$ progenitor $-$ carried out to more than one second postbounce $-$ is an outlier, with the accretion rate plummeting at $\sim$225 ms. The 60-M$_{\odot}$ progenitor loses most of its mass to winds early on, and behaves effectively like a lower mass progenitor. As will be discussed later, the low-mass 9-M${_\odot}$ progenitor explodes more spherically (\citealt{burrows_2019}) than the more massive models studied, and for all intents and purposes, the 9-M$_{\odot}$ progenitor evolution is complete. We see a weak correlation between progenitor mass, accretion rates, and neutrino luminosities. The 25-M$_{\odot}$ progenitor is a clear example of this, with neutrino luminosities roughly 25\% higher than for the 19-$M_{\odot}$ progenitor. The 60-M$_{\odot}$ progenitor is an outlier, with lower luminosities and accretion rates than the 19- and 25-M$_{\odot}$ progenitors. \footnote{We note that in general the mean neutrino luminosity in 2D evinces greater temporal variation than in 3D.} The sharp drop in the accretion rate (Fig.\,\ref{fig:mdot}) and corresponding sharp drop $-$ by as much as 30\% $-$ in the electron-neutrino and anti-neutrino luminosities (Fig.\,\ref{fig:lum}), driven by changes in mass accretion rate, correspond to the accretion of the Si/O interface. The heavy neutrino luminosity is more sensitive to PNS convection (\citealt{radice2017b}) and shows a more muted drop. 

\begin{figure}
    \centering
    \includegraphics[width=0.5\textwidth]{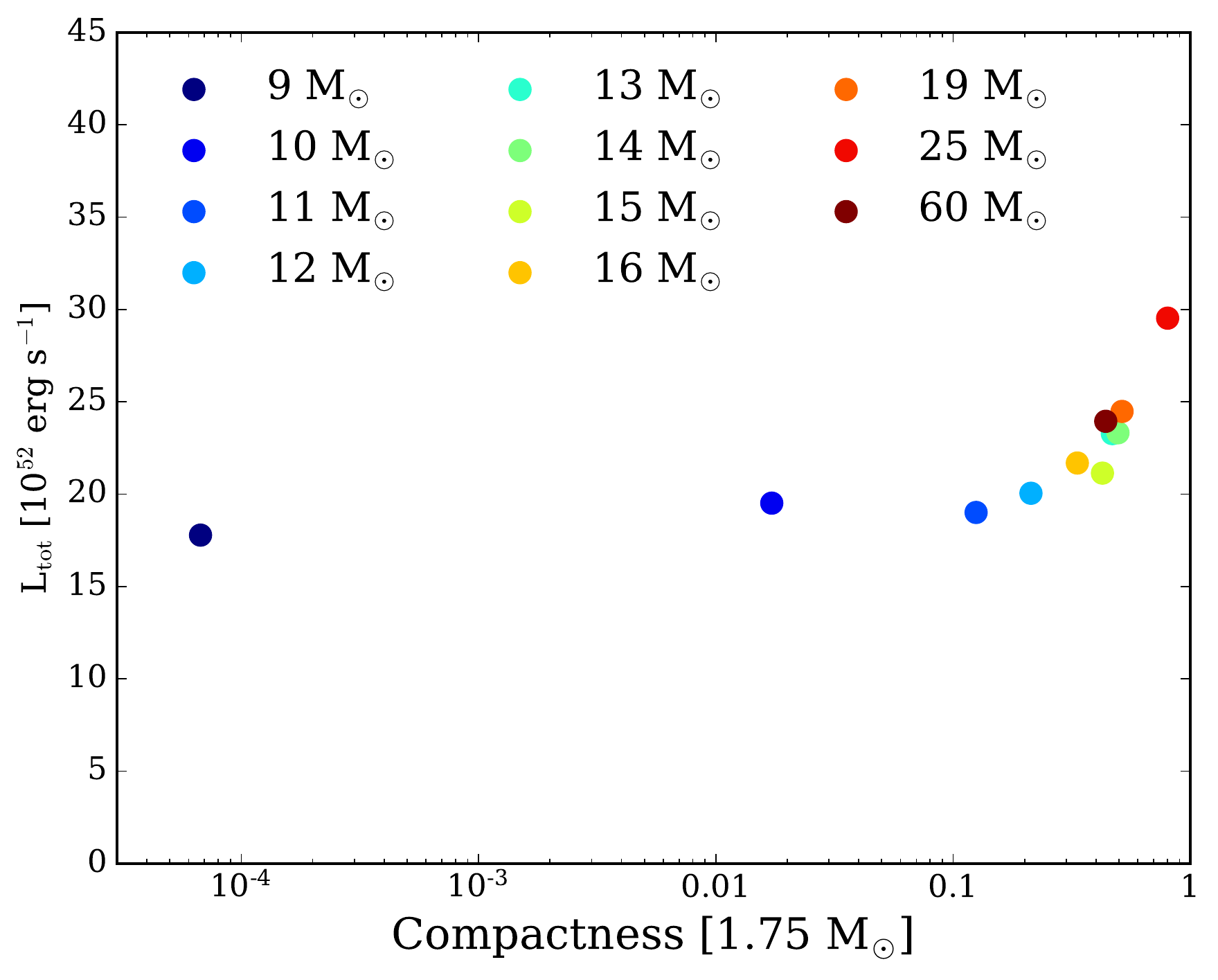}
    \caption{Total neutrino luminosity (in 10$^{52}$ erg s$^{-1}$) summed over all species measured at the post-breakout bump (see Fig.\,\ref{fig:lum}), between 70 $-$ 100 ms, for the various progenitors as a function of the compactness at 1.75 M$_{\odot}$. We note the linear trend towards higher peak luminosity with increasing compactness. The non-exploding models (13-, 14-, 15-M$_{\odot}$) lie slightly below the trend, indicating a lower luminosity for a given compactness. Compactness is just one measure of profile shallowness, which is a multi-dimensional quantity for which fits do not yet exist. We observe explosions for models with higher and lower compactness than the non-exploding 13-, 14-, 15-M$_{\odot}$ progenitors, and emphasize that compactness is not a criterion of explodability.}
    \label{fig:compact}
\end{figure}

The 13-, 14-, and 15-M$_{\odot}$ progenitors lack a sharp Si/O interface (Fig.\,\ref{fig:rho_M}), and fail to explode. The 9-M$_{\odot}$ progenitor also lacks a sharp interface, but explodes by virtue of its steep density profile and low gravitational binding energy. We emphasize that compactness does not correlate with explodability, as models with both higher and lower compactness than the 13-, 14-, and 15-M$_{\odot}$ progenitors do explode, consistent with conclusions from earlier work in 2D (\citealt{vartanyan2018a,radice2017b}) and 3D (\citealt{vartanyan2018b,burrows_2019}). Compactness is just one measure of profile shallowness, which is a multi-dimensional quantity for which useful fits do not yet exist. Compactness does roughly correlate with peak neutrino luminosity (see Fig.\,\ref{fig:lum}) consistent with earlier studies (\citealt{horiuchi2017}) and with mean accretion rate (see Fig.\,\ref{fig:mdot}) for all the exploding models. In Fig.\,\ref{fig:compact}, we plot compactness at 1.75 M$_{\odot}$ against total neutrino luminosity measured at the post-breakout bump. We see that the non-exploding models are outliers and have a slightly smaller neutrino luminosity for a given compactness, and do at late times not strictly follow the trend of higher compactness with higher neutrino luminosity. 

To explore the angle dependence of the neutrino luminosity, we decompose the lab-frame luminosity up to its monopole, dipole, and quadrupole moments, filtering out spurious higher-order terms that are not handled well with the M1 closure scheme at large radii and low optical depth. We obtain for the luminosity: 

\begin{equation}
\begin{split}
 & L_{\nu_i}(\theta,\phi,t)=\, A_0\,Y_{00} + \sum\limits_{m=-1}^1 A_{1m}\,Y_{1m}(\theta,\phi) +\sum\limits_{m=-2}^2 A_{2m}\,Y_{2m}(\theta,\phi)\,,\\
&\mathrm{where}\\
&A_{ij}(t) =  \int_{\Omega} r^2 F_r [t,r,\theta,\phi] \times Y_{ij} [\theta,\phi]\, d\Omega\,,
\end{split}
\end{equation} and where the terms on the right-hand-side corresponding to the monopole, dipole, and quadrupole terms, respectively, for each neutrino species i $\in$ \{$\nu_e$,\,$\bar{\nu}_e$,\,$\nu_{\mu}$\}. $F_r$ is the radial flux outwards at a given radius, here taken to be 250 km.

\begin{figure*}
    \centering
    \includegraphics[width=0.99\textwidth]{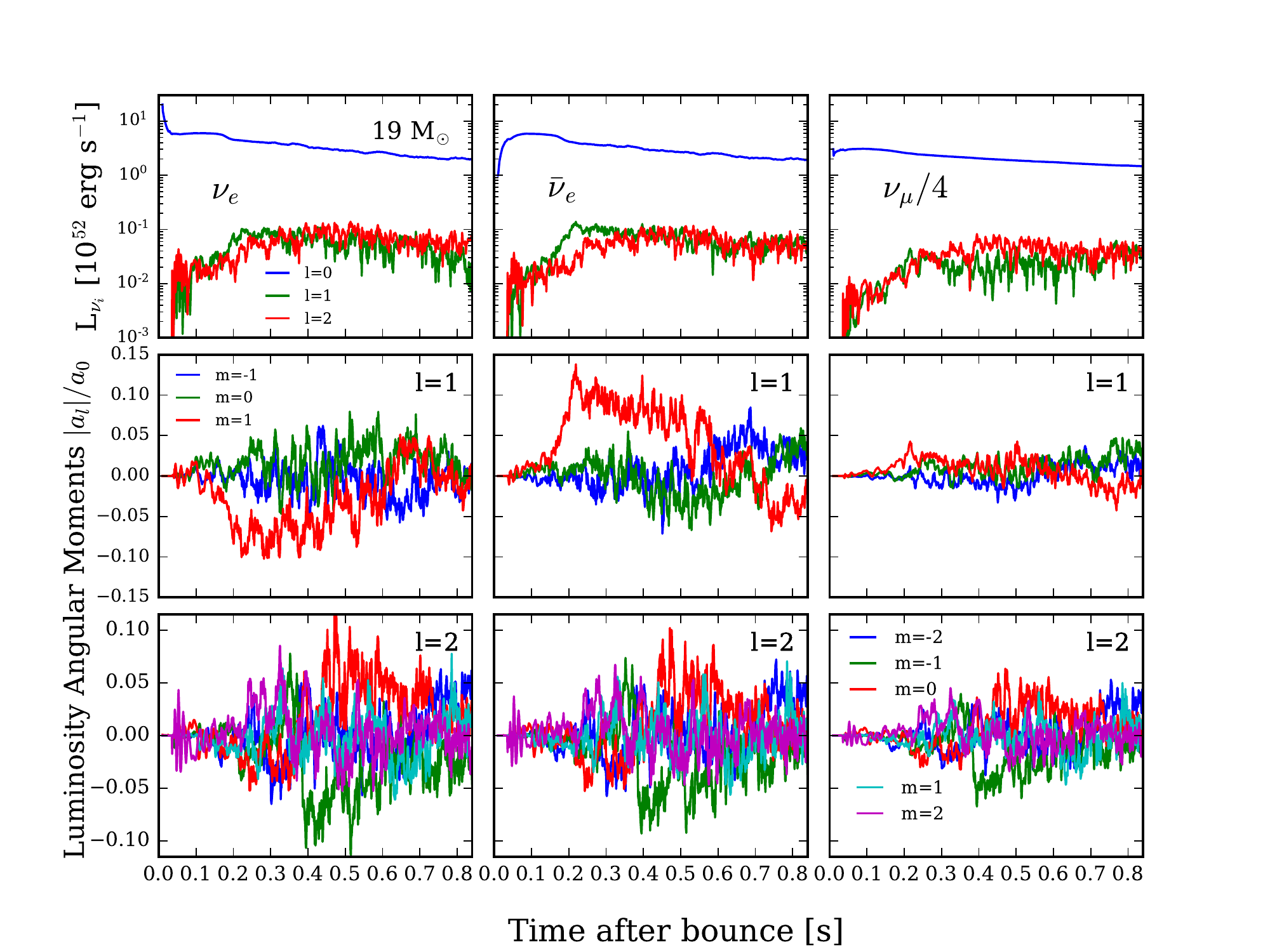}

    \caption{The luminosity decomposition by angle for the 19-M$_{\odot}$ progenitor for the $\nu_e$, $\bar{\nu}_e$, and $\nu_\mu$ species as a function of time after bounce (in seconds), truncated to the quadrupole term. In the top set of panels, we sum in quadrature over the $m$s. The dipole and quadrupole terms are never greater than several percent of the monopole term, and the heavy-neutrinos have the smallest angular deviations from spherical asymmetry. Fractional asymmetries in the neutrino luminosity are much smaller than fractional asymmetries in the shock radius. The middle, bottom horizontal panels provide the corresponding normalized angular moments of the neutrino luminosity for all the $m$ components for $\ell=1\,,\ell=2$, respectively.}
        \label{fig:lum_lm}

\end{figure*}

In Fig.\,\ref{fig:lum_lm}, we plot the luminosity decomposition in angle for the 19-M$_{\odot}$ progenitor for the $\nu_e, \bar{\nu}_e,$ and $\nu_\mu$ neutrino species as a function of time after bounce, truncated at the quadrupole term. In the top panel, we plot the monopole, dipole, and quadrupole terms for each species, summed in quadrature over azimuthal moments $m$, and in the bottom two panels we plot the dipole and quadrupole terms for each moment $m$. The dipole and quadrupole terms are never greater than several percent of the monopole term, and the heavy-neutrinos have the smallest deviations from spherical asymmetry. We see the dipole undergo a small number phases of growth until $\sim$200 ms, where turbulence becomes significant and we see a steep rise in the luminosity dipole. For all our models, the dipole component of the neutrino luminosity for all species is never more than $\sim$5-6\% of the angle-averaged luminosity. We note that angle asymmetries in the luminosity are much smaller than asymmetries in the shock radius. 

\begin{figure*}
    \centering
    \includegraphics[width=0.49\textwidth]{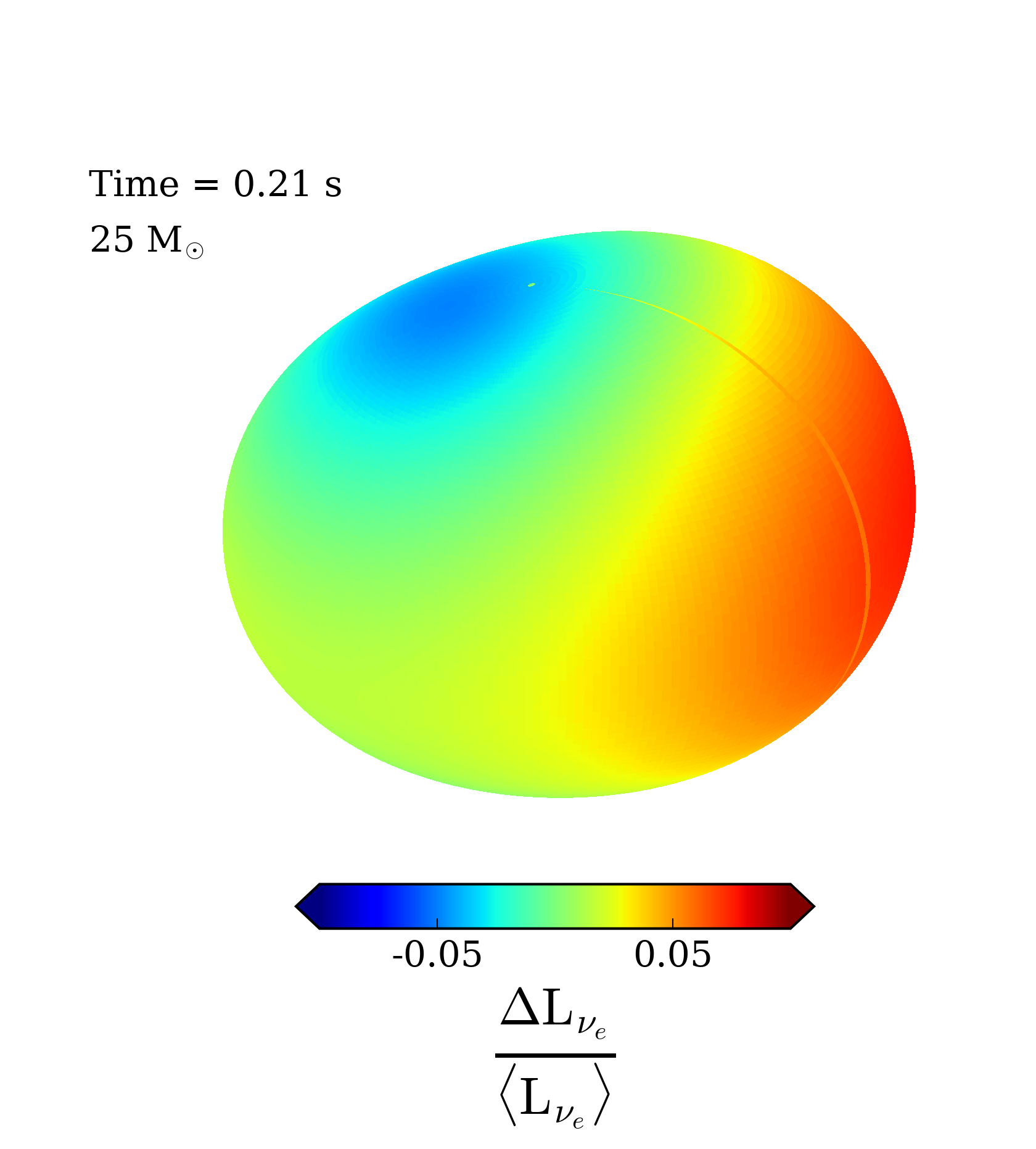}
    \hfill
    \includegraphics[width=0.49\textwidth]{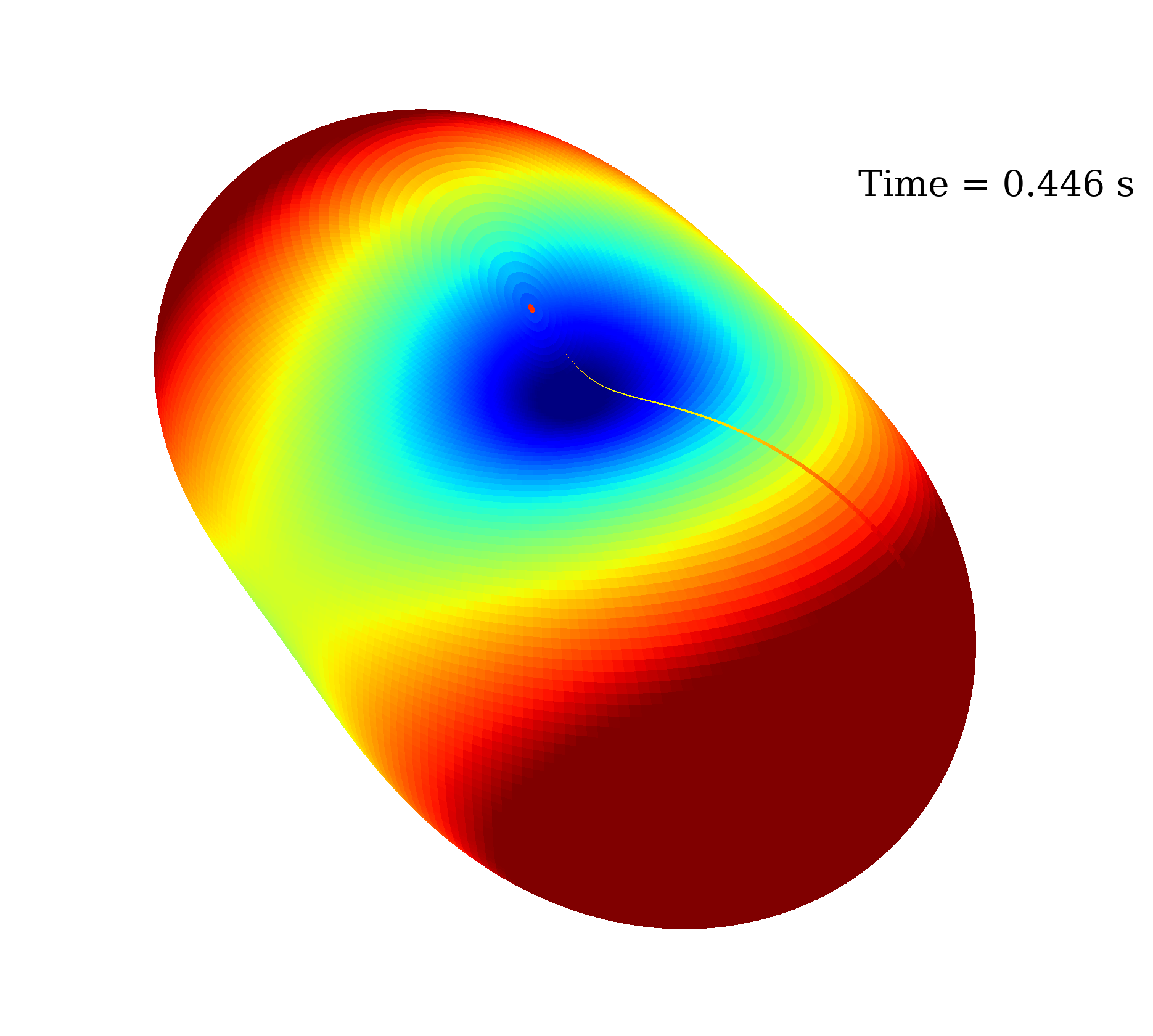}
    \includegraphics[width=0.49\textwidth]{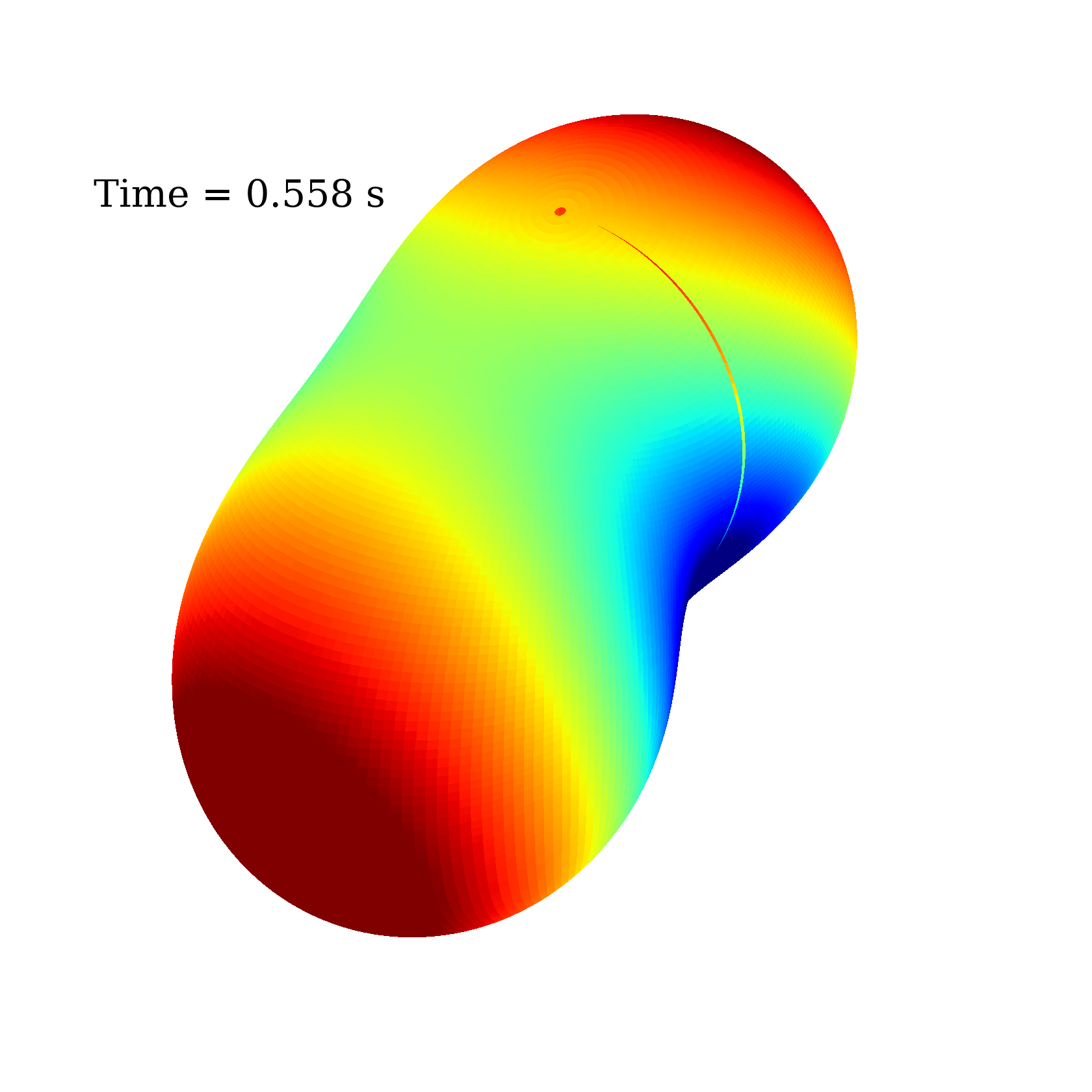}
    \includegraphics[width=0.49\textwidth]{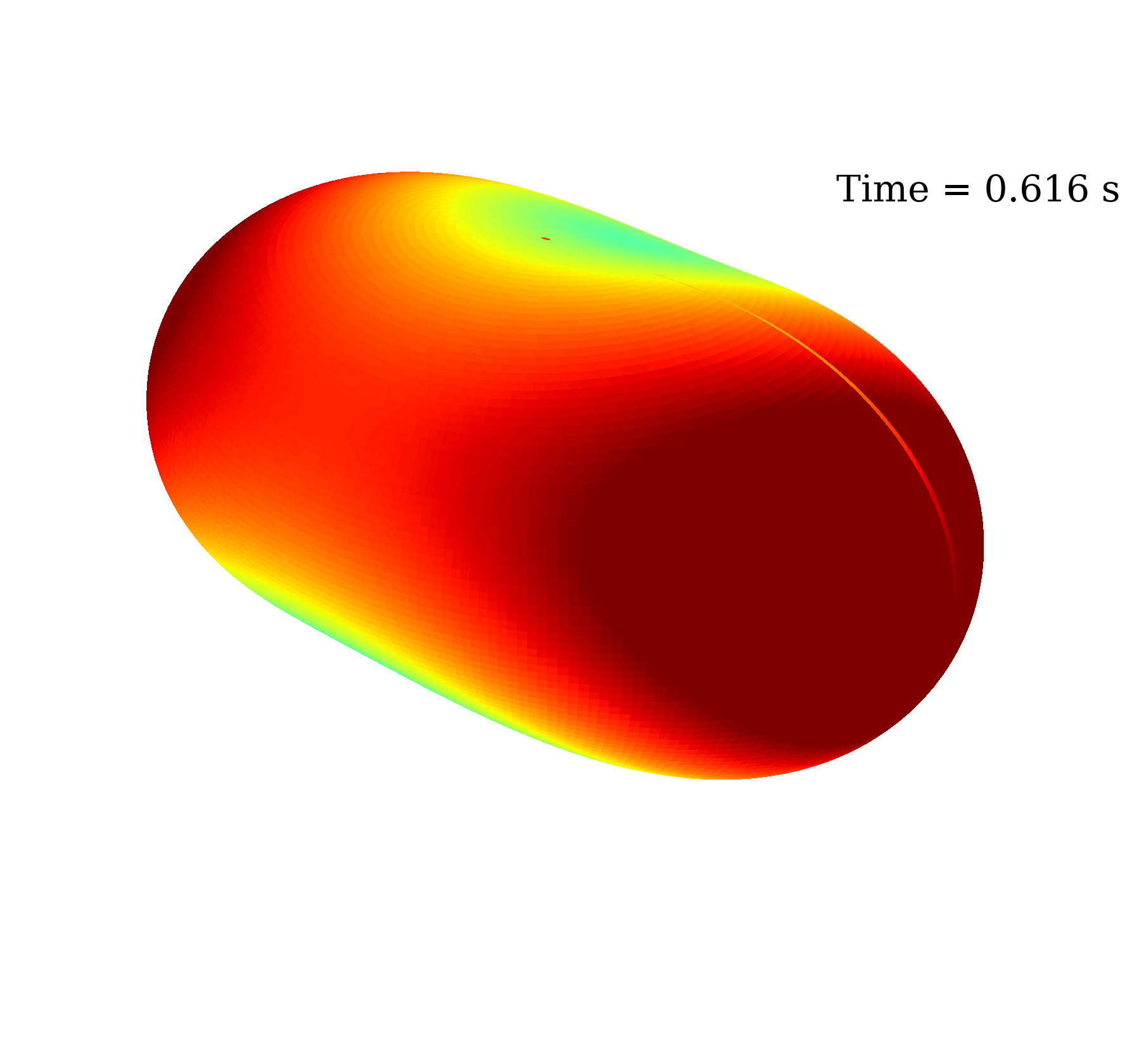}
    \hfill
    \caption{Fractional deviation from the mean in the $\nu_e$ luminosity as a function of viewing angle and at various times for the 25-M$_{\odot}$ progenitor. In the top of each panel. We color-code a region comensurate with the magnitude of the bulge, with color and radius are redundant. Cool-colored dimples indicate lower-than-average neutrino luminosities, and warm-colored protrusions higher-than-average. The stripe is the vestigial axis artifact.}
    \label{fig:lum_25contour}
\end{figure*}

In Fig.\,\ref{fig:lum_25contour}, we plot in 3D the fractional variation in the electron-neutrino luminosity as a function of viewing direction and at various times for the 25-M$_{\odot}$ progenitor. In the top of each panel, we color-code and contour the fractional variation, with color and contour redundant with each other. Cool-colored dimples indicate lower-than-average neutrino luminosities, and warm-colored protrusions indicate higher-than-average luminosities. The narrow stripe is the vestigial axis artifact. We see large-scale structure of the neutrino luminosity, with typical variations over viewing direction that increase with time to 5\%$-$10\%.

\begin{figure}
    \centering
    \rotatebox[origin=c]{-90}{%
    \begin{minipage}{1.2\textwidth}
    \centering
    \includegraphics[width=0.49\textwidth]{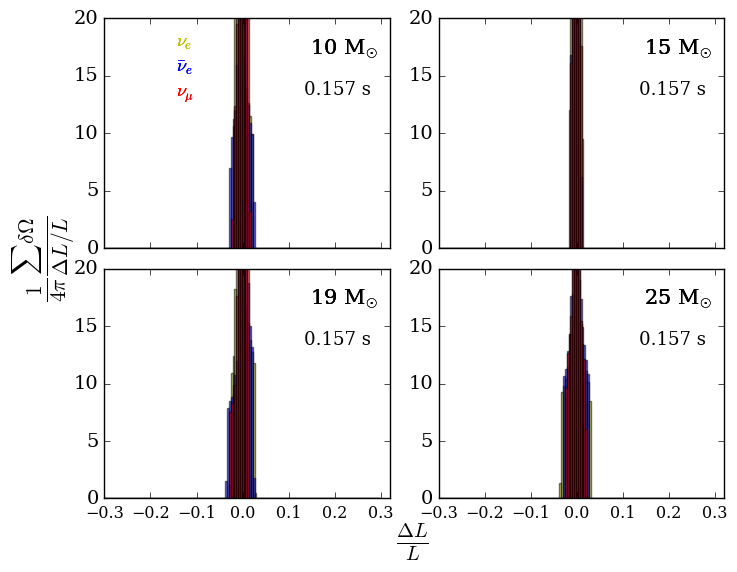}
    \hfill
    \includegraphics[width=0.49\textwidth]{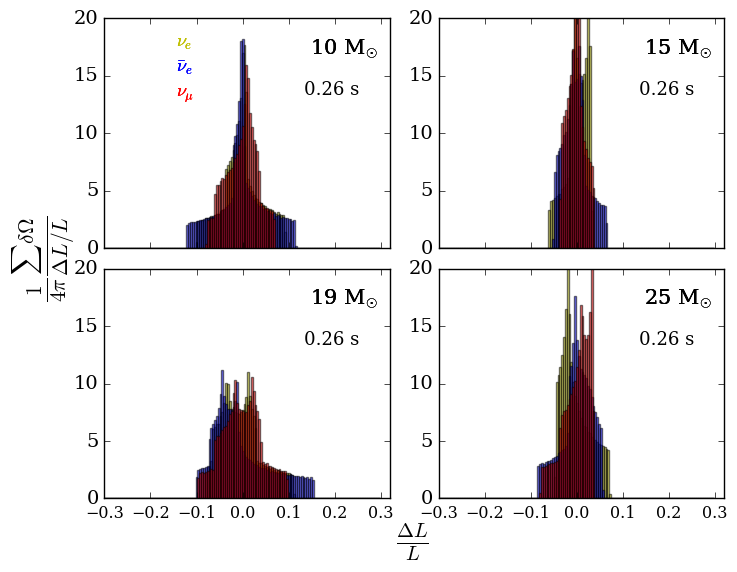}
    \includegraphics[width=0.49\textwidth]{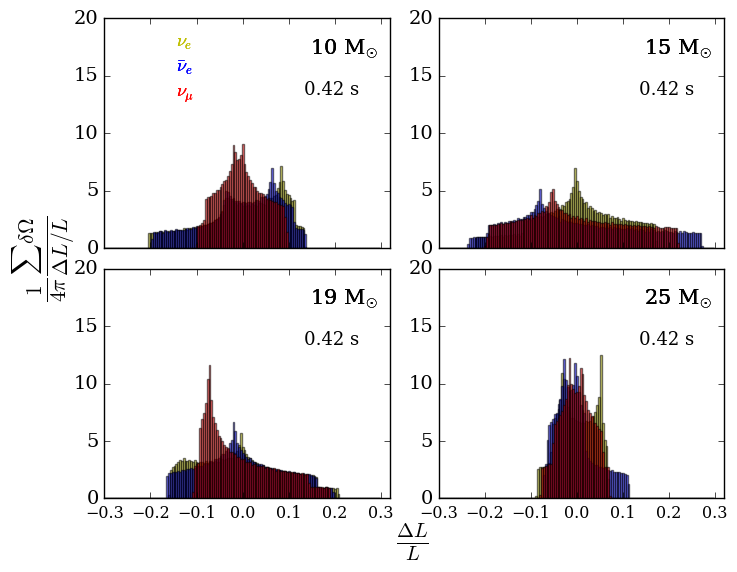}
    \includegraphics[width=0.49\textwidth]{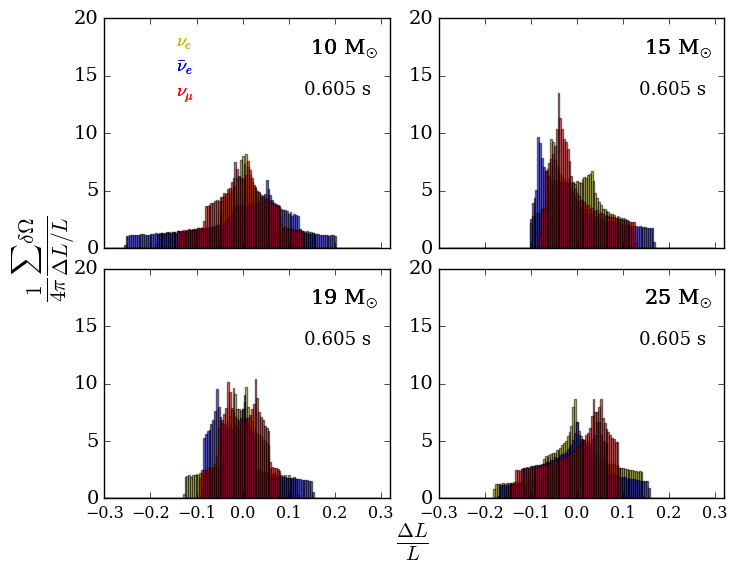}
    \hfill
    \caption{Histograms of the distributions of fractional deviation of the neutrino luminosity (of the three different species, yellow: $\nu_e$, blue:$\bar{\nu}_e$, red: $\nu_\mu$) at 250 km  from the mean ($\frac{\Delta L}{L}$) radiated into the 4$\pi$ steradians of the sky for several different progenitors. The integral under each histogram curve is normalized to one. All models begin with isotropic neutrino emission then evolve towards larger variations with viewing angle. Instantaneous neutrino emission can vary by as much as 40\% with viewing direction. We note that heavy-neutrinos typically show less temporal variation in luminosity. The 15-M$_{\odot}$ progenitor, which does not explode, is consistently more isotropic in neutrino emission, even at later times.}
        \label{fig:dNdL_hist}
\end{minipage}
}
\end{figure}

In Fig.\,\ref{fig:dNdL_hist}, we plot histograms of the fractional emitting area at 250 km of the deviation from the mean neutrino luminosity for the three different species for several different progenitors. All models begin with isotropic neutrino emission and then evolve towards larger variations by viewing angle. Instantaneous neutrino emission can vary by as much as 40\% over this sphere. We note that heavy-neutrinos typically show less angular variation in luminosity. The 15-M$_{\odot}$ progenitor, which does not explode, is consistently more isotropic in neutrino emission, even at later times.

\begin{figure*}
    \centering
     \includegraphics[width=1.04\textwidth]{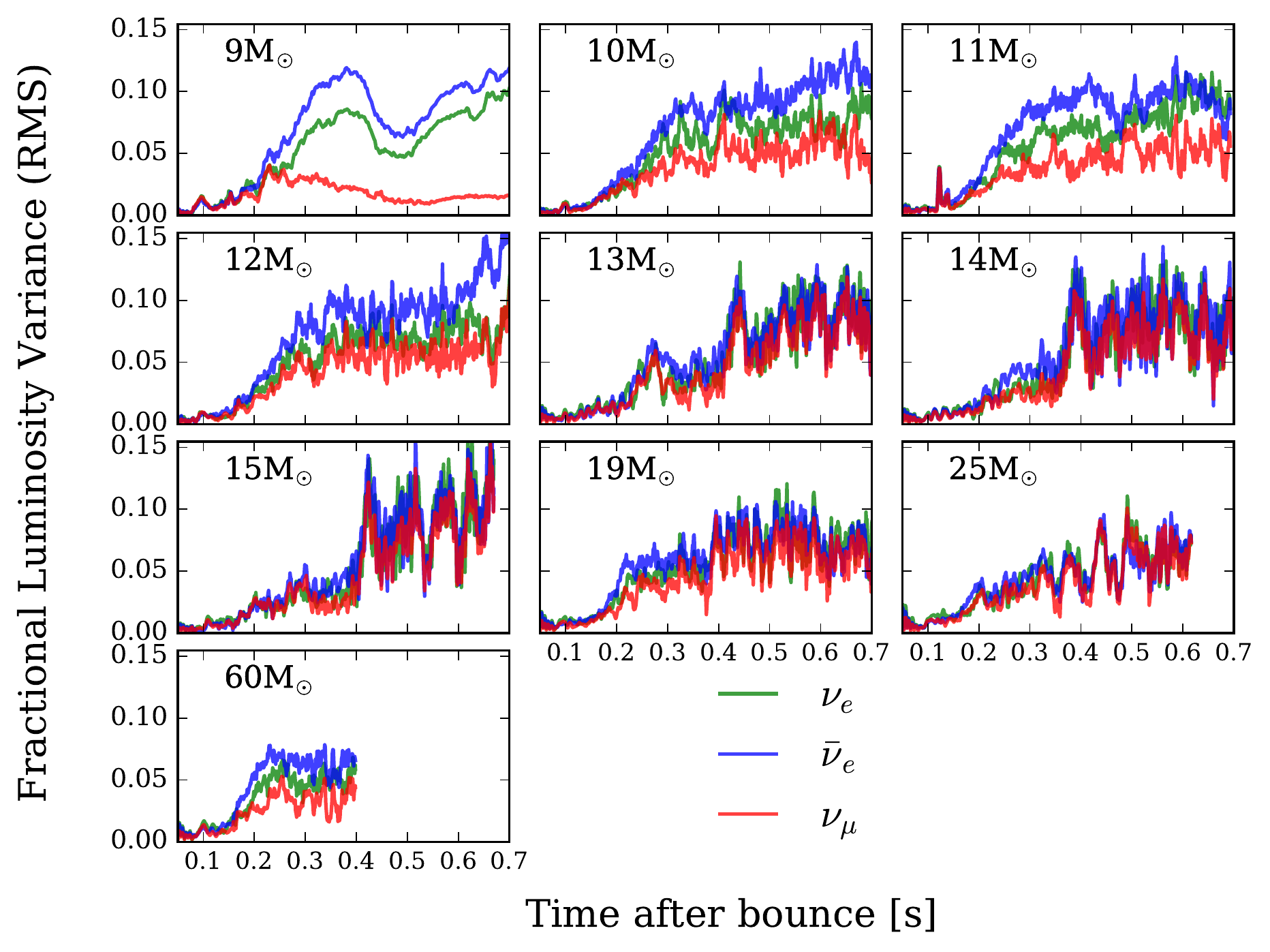}

    \caption{Root mean square (RMS) variation around the mean of the neutrino luminosities for the different neutrino species, normalized by the average neutrino luminosity for that species, as a function of time after bounce (in seconds). The fractional RMS is typically less than 8\% of the neutrino luminosity, even at late times. Note the remarkable similarity in behavior for the different species. We observe a hierarchy in the fractional RMS, with the electron anti-neutrinos showing the greatest deviation from the mean, and the heavy-neutrinos the least. The 9-M$_{\odot}$ progenitor is the only progenitor that shows a drop in the RMS in the heavy-neutrino species just after $\sim$200 ms. This corresponds to the truncation of accretion and the end of the dynamic evolution of the 9-M$_{\odot}$ supernova - this model has asymptoted in explosion energy and all relevant diagnostics.}
        \label{fig:lum_rms}
\end{figure*}

We plot the corresponding solid-angle averaged RMS neutrino luminosity about its mean in Fig.\,\ref{fig:lum_rms} for the 19-M$_{\odot}$ progenitor. We see a hierarchy in fractional RMS neutrino luminosity by species; electron anti-neutrinos show the most variation, and heavy-neutrinos the least. The RMS variation increases with time; however, even at late times, the RMS variation is just $\sim$8\%. Immediately after breakout, the variations are much smaller, roughly $\sim$1\%. We see remarkable correlation between the different neutrino species in their RMS variation. This trend holds for all eleven models included in this study. 

Given the small angular variation by direction in neutrino luminosities early on, future neutrino detections will not depend much on viewing angle after breakout and can be used to differentiate between core density structure and compactness (Fig.\,\ref{fig:lum}). Even up to 200$-$300 ms, variation by viewing angle will be dwarfed by the intrinsic differences in the luminosity by progenitor. Furthermore, observation of the $\nu_e$ and $\bar{\nu}_e$ neutrino luminosity can identify the presence and time of Si/O accretion.

\subsection{Neutrino Time Series}\label{sec:timeseries}

\begin{figure*}
    \centering
    \includegraphics[width=0.49\textwidth]{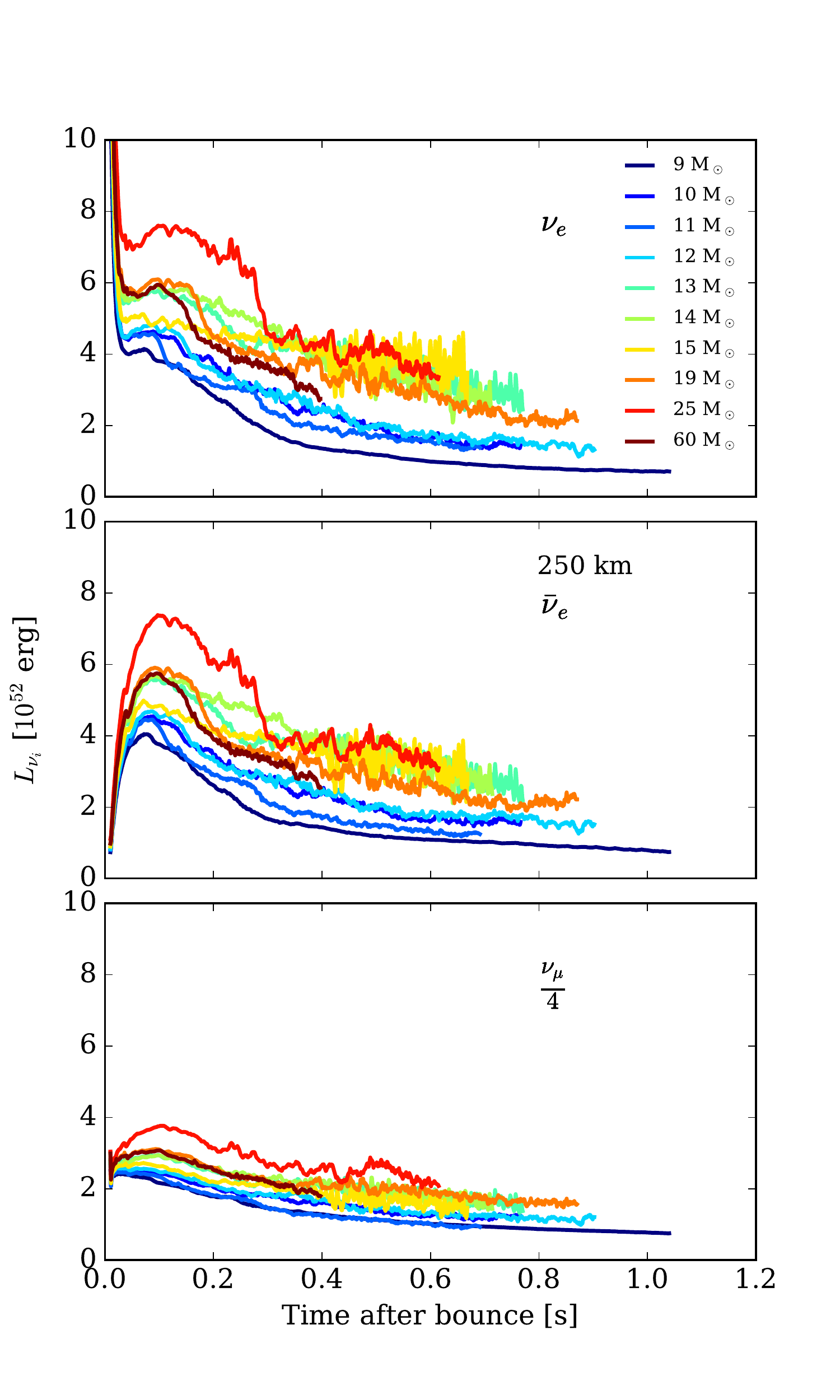}
    \includegraphics[width=0.49\textwidth]{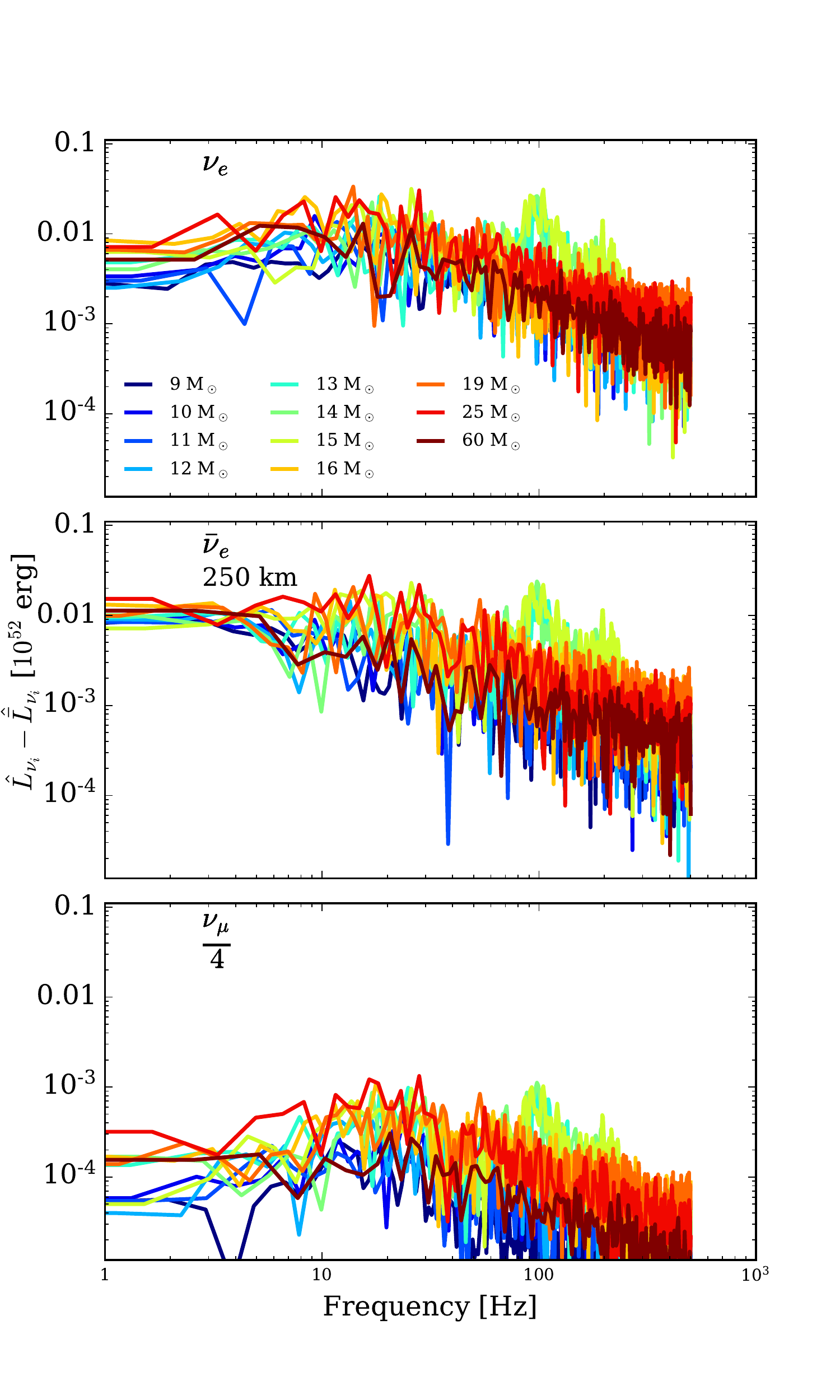}
    \caption{\textbf{Left}: The neutrino luminosity (in 10$^{52}$ erg s$^{-1}$) as a function of time after bounce (in seconds) for the three species along an arbitrarily chosen viewing direction (selected as $\theta = 49^{\circ}\,,\phi = 91^{\circ}$ in the spherical coordinate system of the supernova, compare to the angle-average neutrino luminosity in Fig.\,\ref{fig:lum}). Note the greater variability for the non-exploding models. \textbf{Right}: The Fourier transform of the neutrino luminosities (in 10$^{52}$ erg), subtracting out the running average over 30 ms, for the three species as a function of time after bounce (in seconds).}
        \label{fig:lum_series}
\end{figure*}

Here, we explore the time variability along an arbitrarily chosen viewing direction of the neutrino luminosity for the various species. We arbitrarily choose a viewing direction along $\theta = 49^{\circ}\,,\phi = 91^{\circ}$ in the spherical coordinate system of our simulated supernova. In the left panel of Fig.\,\ref{fig:lum_series}, we plot the neutrino luminosity of the different species as a function of time after bounce along this viewing direction (to be compared with Fig.\,\ref{fig:lum}, where we plot the angle-averaged neutrino luminosities). To probe the temporal variation, we subtract out the luminosity running average along the same viewing direction, using a window of 30 ms. We see a rough trend with progenitor mass of the neutrino temporal variations, with the 9-M$_{\odot}$ progenitor having less than 1\% variations from the mean with time. The non-exploding models, 13-, 14-, and 15-M$_{\odot}$, show the greatest variation of all the models after 200 ms, with the development of the SASI manifesting as high-amplitude, high-frequency variations in the luminosity. These models show average temporal variation of $\sim$8\%, with variations as high as 25\%. By contrast, the exploding models show average temporal variation over 2$-$4\%, with the 16-, 19-, 25-M$_{\odot}$ progenitors exhibiting variations as high as 10\%.

Generally, the $\nu_e$ and $\bar{\nu}_e$ neutrino luminosities show the greatest temporal (and spatial) variation, while the $\nu_\mu$-neutrino luminosity exhibits the least temporal and spatial variation. In the right panel of Fig.\,\ref{fig:lum_series}, we plot the Fourier transform of the neutrino luminosity along the same viewing direction, subtracting out the running average along this same direction. The high-frequency components are associated with $\sim$few millisecond timescale PNS convection. The peaks in the heavier progenitors, around $\sim$10 Hz, indicate large-scale explosion asymmetries. Note the 100 Hz peak in the non-exploding models, indicative of SASI, is absent in the exploding models.

\section{Temporal and Directional Variations of Gravitational Wave Emissions}\label{sec:GW}

In this section, we explore the spatial and temporal variations of the gravitational wave signatures of our suite of progenitors. We provide gravitational wave data for all models except the 16-M$_{\odot}$, for which we did not calculate the gravitational quadrupole moments. We follow \cite{oohara97} and \cite{andresen2017}, with the gravitational strain polarizations defined as 

\begin{subequations}
\begin{align}
    h_+ &= \frac{1}{r}\left(\ddot{Q}_{\hat\theta\hat\theta} - \ddot{Q}_{\hat\phi\hat\phi}\right)\\
    h_\times &= \frac{2}{r}\ddot{Q}_{\hat\theta\hat\phi}
\end{align}
\end{subequations} and the quadrupole moments defined as 

\begin{subequations}
\begin{align}
\begin{split}
Q_{\hat\theta\hat\theta} &= (Q_{xx} \mathrm{cos}^2 \phi + Q_{yy} \mathrm{sin}^2 \phi + 2Q_{xy} \mathrm{sin}\,\phi \, \mathrm{cos}\,\phi)\, \mathrm{cos}^2 \theta 
\\&+ Q_{zz} \mathrm{sin}^2 \theta - 2\left(Q_{xz} \mathrm{cos}\, \phi + Q_{yz} \mathrm{sin} \,\phi \right) \mathrm{sin}\, \theta\, \mathrm{cos}\, \theta
\end{split}\\
Q_{\hat\phi\hat\phi} &= Q_{xx}\mathrm{sin}^2 \phi + Q_{yy}\mathrm{cos}^2\phi - 2Q_{xy}\mathrm{sin}\,\phi\,\mathrm{cos}\,\phi\\
\begin{split}
Q_{\hat\theta\hat\phi} &= (Q_{yy} - Q_{xx})\,\mathrm{cos}\,\theta\,\mathrm{sin}\,\phi\,\mathrm{cos}\,\phi \\&+ Q_{xy}\mathrm{cos}\,\theta\,(\mathrm{cos}^2\phi - \mathrm{sin}^2\phi) + Q_{xz}\mathrm{sin}\,\theta\,\mathrm{cos}\,\phi \, .
\end{split}
\end{align}
\end{subequations}

\begin{figure*}
    \centering
    \includegraphics[width=.8\textwidth]{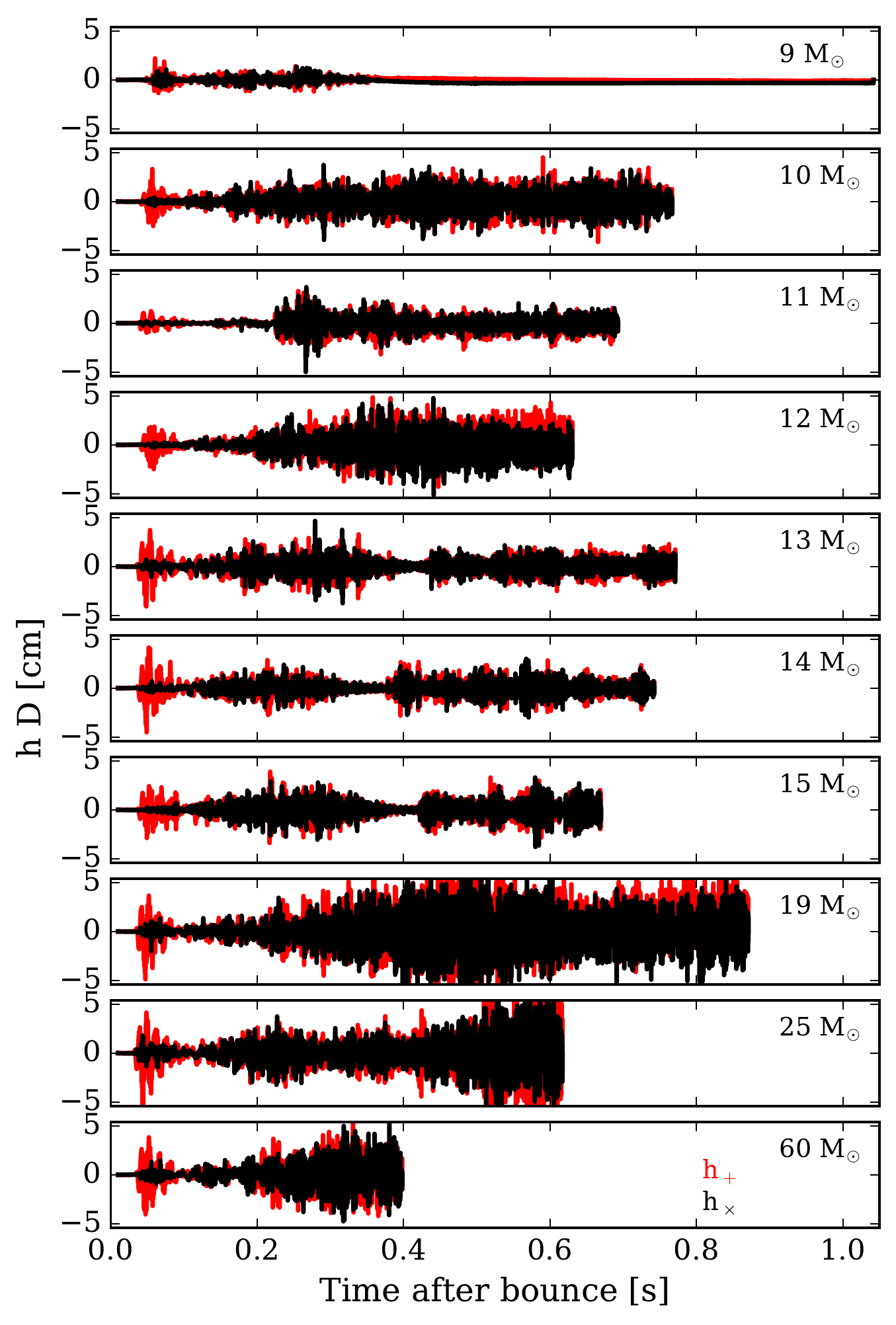}
    \caption{Gravitational wave strain h$_{+,\times}$D (in centimeters, where D is the distance) as viewed along the x-axis in the coordinate system of the supernova  as a function of time after bounce (in seconds) for the various models. The red lines show h$_+$ and the black lines h$_\times$. The strong prompt h$_+$ strain for the selected viewing angle, absent in h$_\times$, corresponds to prompt convection and is indicative of the symmetry of the perturbations implemented. Afterwards, the two polarizations roughly follow each other in evolution. Note the cessation in the GW signal for both polarizations for the 9-M$_{\odot}$ progenitor, shortly after accretion ends (\protect\citealt{radice2019}). For the remaining models, the strain ramps up within $\sim$200 ms, and its magnitude is approximately correlated with progenitor mass. For the non-exploding models (13-, 14-, and 15-M$_{\odot}$ progenitors), the accretion rate, after growing for $\sim$200 ms, is `pinched' and drops until $\sim$400 ms postbounce (where we see the spiral SASI develop), when it is revitalized.}   \label{fig:gw_strain}
\end{figure*}

In Fig.\,\ref{fig:gw_strain}, we plot the gravitational wave strain times a distance (D) as viewed along the x-axis in the coordinate system of the supernova for the 10 models (neglecting the 16-M$_{\odot}$ model) and including both linear polarizations h$_+$ and h$_\times$. The strong prompt signal in h$_+$ (and absent in h$_\times$) in all models for the chosen viewing angle corresponds to the onset of prompt convection and indicates the symmetry of the imposed perturbations.  The strain ramps up within the first $\sim$200 ms, and subsequent to the prompt convection phase, both polarizations roughly follow each other in evolution. We see `packets' in the strain lasting $\sim$50 ms corresponding to episodic accretion.  Furthermore, in contrast with \citealt{pajkos2019}, we find that the gravitational wave strain at core bounce, and throughout, generally increases with progenitor mass. We identify for all exploding models a direct correlation between the progenitor accretion rate and the magnitude of the gravitational strain, as concluded in \cite{radice2019}. The 9-, 10-, and 11-M$_{\odot}$ models have the lowest accretion rates, and correspondingly, the smallest strains, while the 19-, 25-, and 60-M$_{\odot}$ progenitors have the highest accretion rates and, correspondingly, the largest strains. For the 9-M$_{\odot}$, in particular, early cessation of the GW signal corresponds to the early cessation of accretion. This emphasizes the importance of carrying out simulations longer in 3D to understand their late-time behavior, through the end of the accretion phase, and the implications for observable signatures.  

We note that the non-exploding 13-, 14-, 15-M$_{\odot}$ models have a weaker strain at later times despite sustained accretion. Furthermore,  the gravitational strain for these non-exploding models grows for $\sim$200 ms, then is `pinched' and decreases until $\sim$400 ms postbounce (when the spiral SASI develops), where it shows renewed growth.

\begin{figure*}
    \centering
    \includegraphics[width=0.43\textwidth]{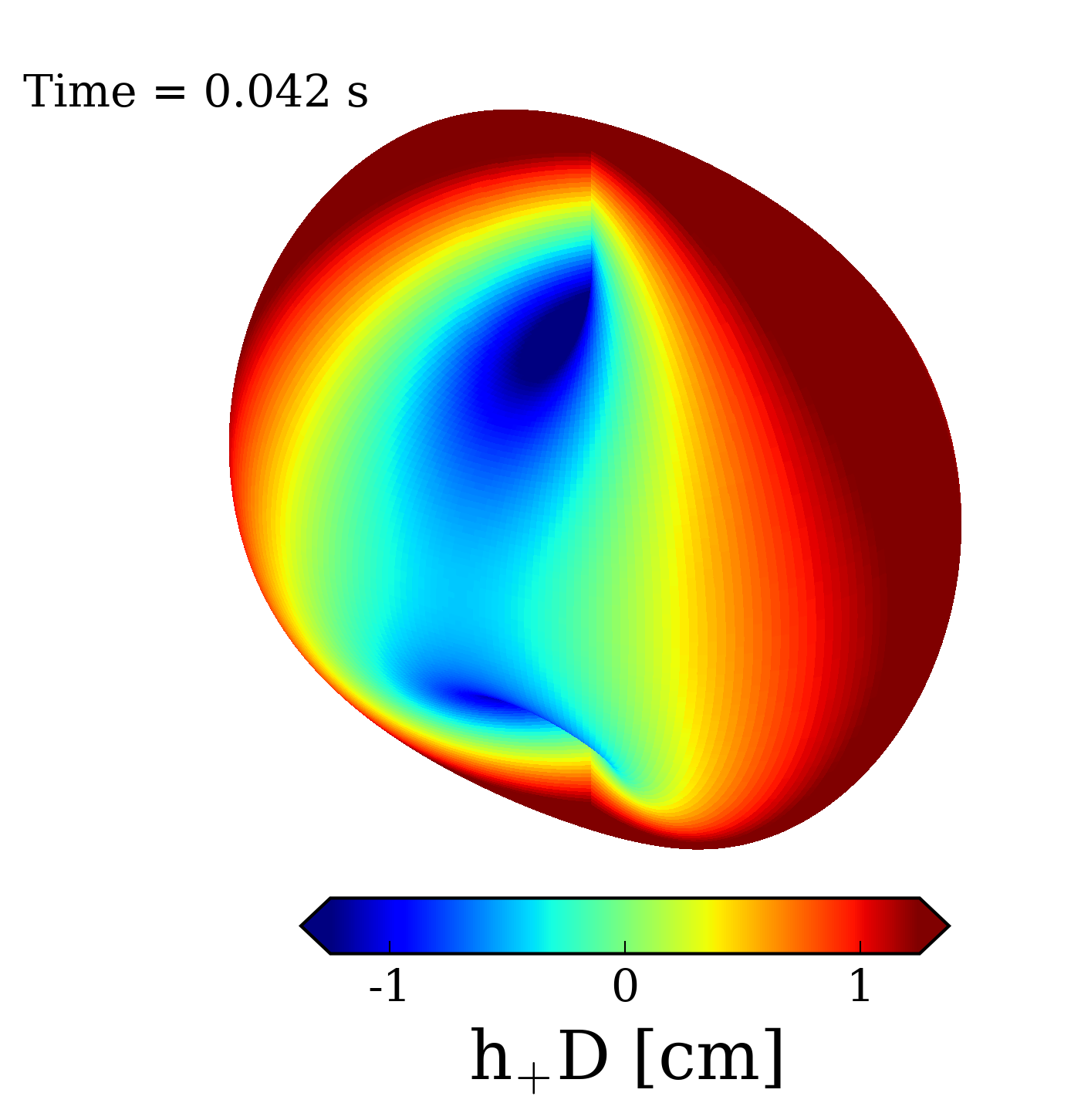}
    \hfill
    \includegraphics[width=0.43\textwidth]{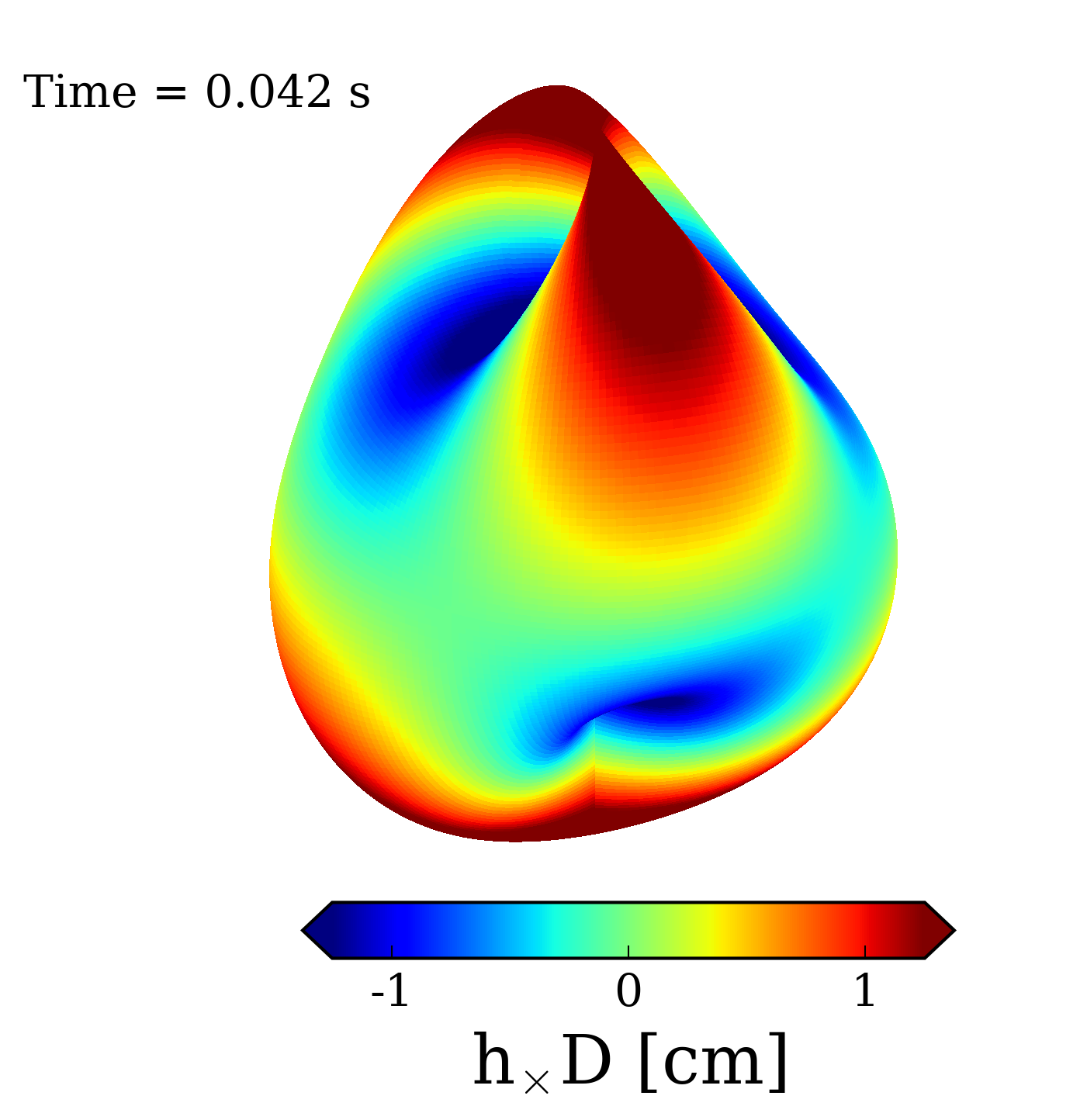}
    \includegraphics[width=0.43\textwidth]{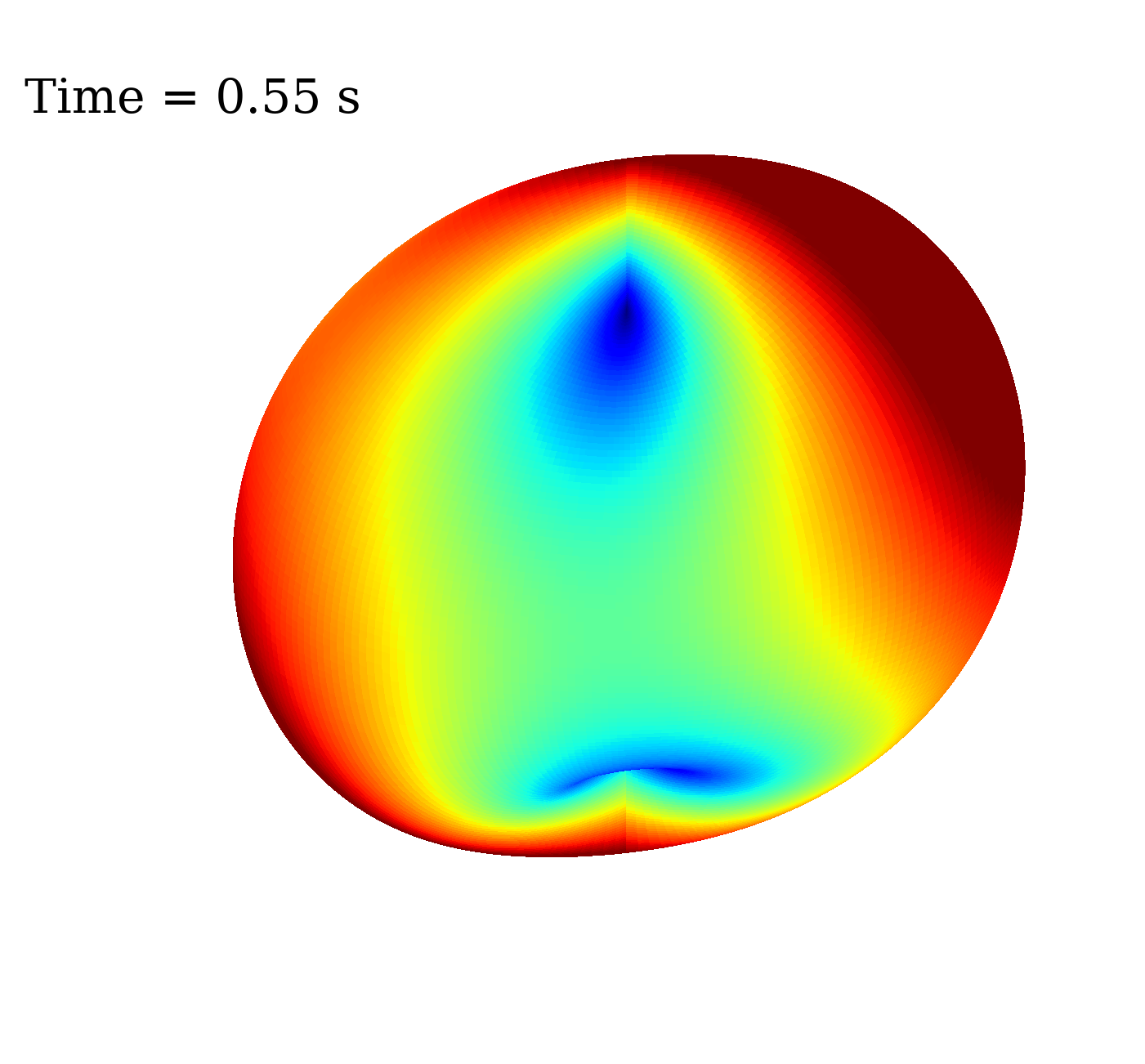}
    \hfill  
    \includegraphics[width=0.43\textwidth]{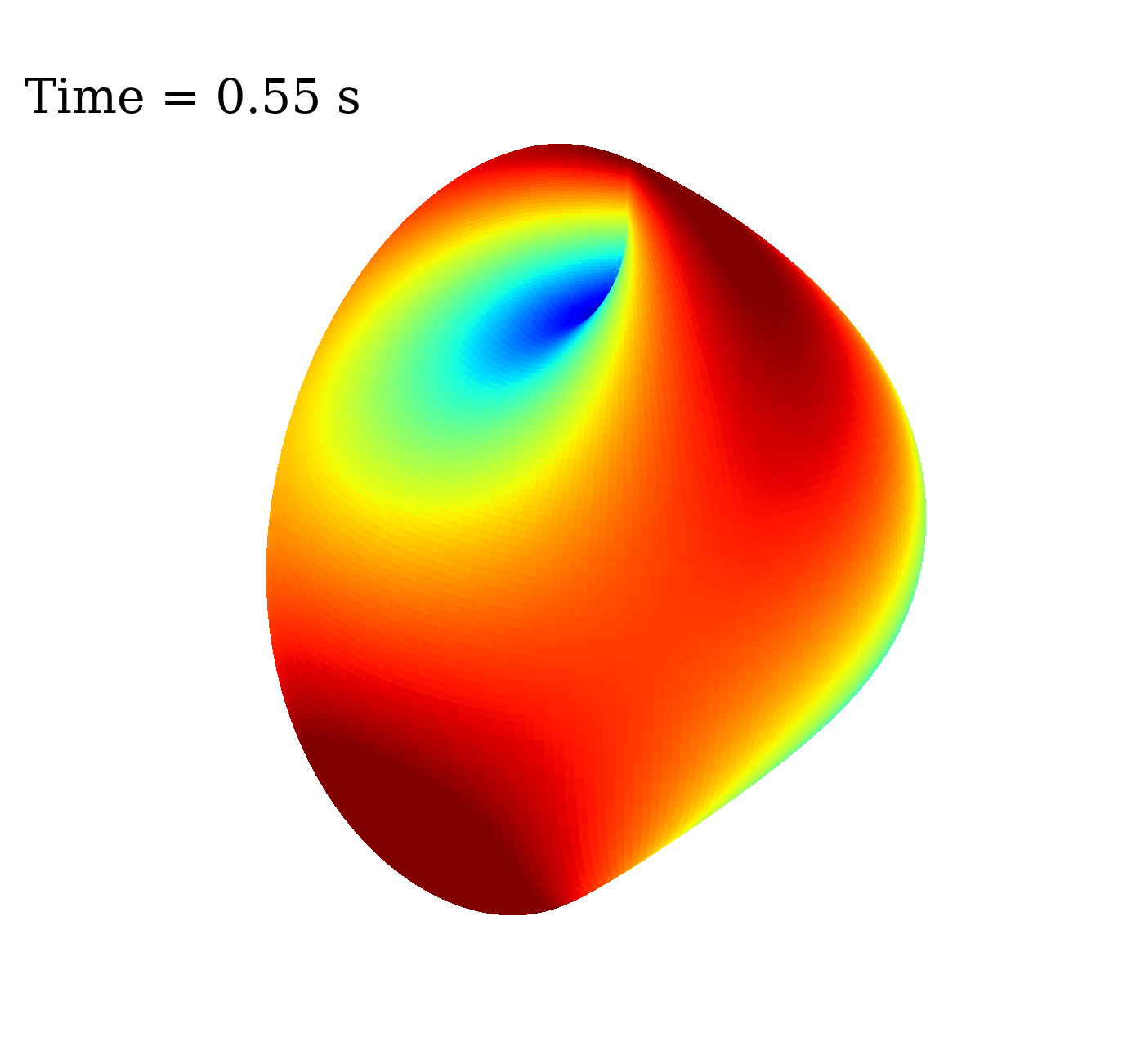}
    \includegraphics[width=0.43\textwidth]{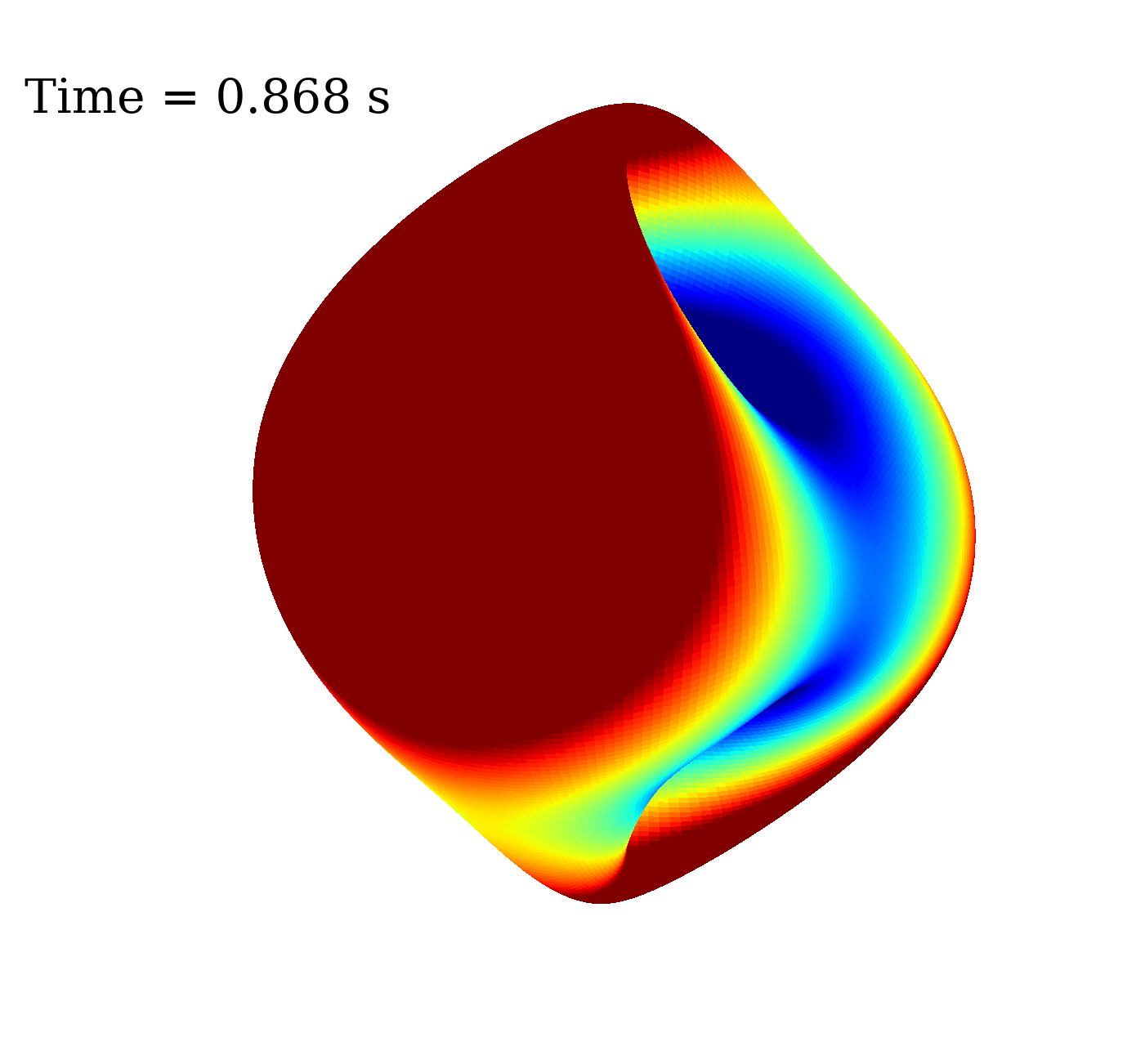}
    \hfill
    \includegraphics[width=0.43\textwidth]{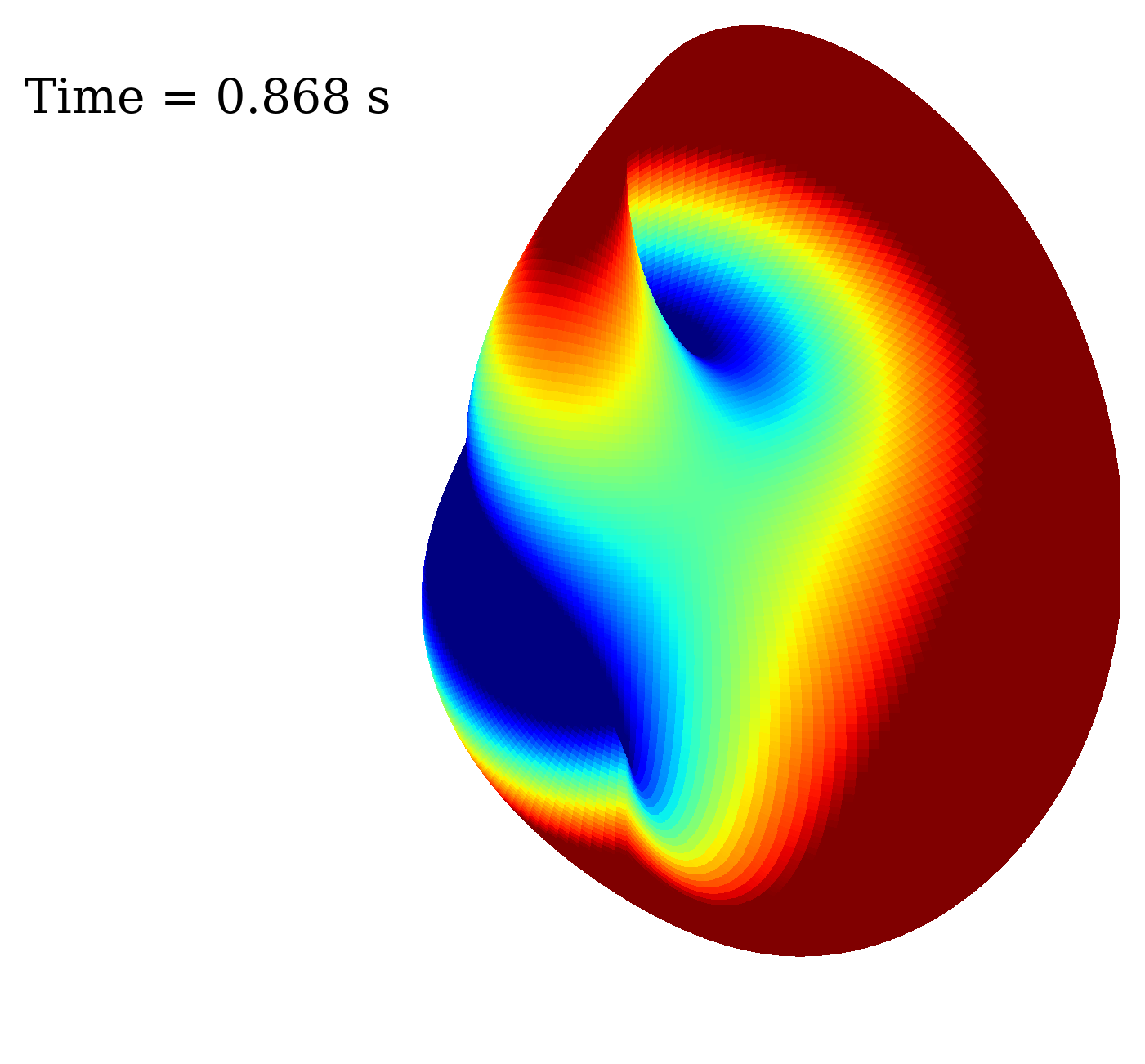}
    \caption{Gravitational strain h$_{+,\times}$ D in cm (h$_+$ \textbf{left}, h$_\times$ \textbf{right}) as a function of viewing angle and at various times for the 19-M$_{\odot}$ progenitor. The contours are overplotted on a surface of 5 cm to accentuate the variations. The color and contours are redundant, with hotter colors and convex surfaces indicating positive strains, and cooler colors and convex surfaces negative strains. We see variations in gravitational wave emission on sub-millisecond timescales associate with $p$-modes in the turbulent region and the frequency growth of the $f$-mode (see Fig.\,\ref{fig:SASI_spec}). The morphologies vary between the two polarizations, with the h$_+$ contour surface shaped like a pinched dumpling, and the h$_\times$ surface a guitar pick.}
        \label{fig:GW_19contour}
\end{figure*}

In Fig.\,\ref{fig:GW_19contour}, we plot the spatial distribution of the gravitational strains h$_+$ D and h$_\times$ D, as dimples on a sphere of radius 5 cm, with the color and magnitude of the dimple corresponding redundantly to the strain. Hot colors and convex surfaces correspond to positive strain; cool colors and concave surfaces correspond to negative strain. The strain varies on sub-millisecond timescales with a dominant, large-scale quadrupolar morphology that differs between the different polarizations.

\section{SASI}\label{sec:SASI}
In this section, we provide an aside on the SASI.
The standing accretion shock instability (SASI; \citealt{foglizzo2002,blondin2003,blondin_shaw,foglizzo2012}), is a vortical-acoustic hydrodynamic instability in the post-shocked region manifested (when it appears) by non-radial oscillating motion. To identify the SASI, we search for a low-frequency (100 $-$ 250 Hz) gravitational wave signature lasting  several hundred milliseconds (\citealt{kuroda2016,andresen2019}). We confirm the presence of the SASI by looking for a stately shock dipole in Fourier space at similar frequencies. We find evidence for the SASI in four of the models considered: the 13-,14-, and 15-M$_{\odot}$ progenitors (which do not explode), and the 25-M$_{\odot}$ progenitor (which explodes later, around 275 ms). All three of the non-exploding models also show an $m=1$ spiral SASI mode (\citealt{blondin2005}) developing $\sim$400 ms postbounce, after the early SASI phase, when the stalled shock radius has receded. The spiral SASI is a three-dimensional feature observed in earlier simulations (\citealt{kuroda2016,summa2018,andresen2019}) and cannot be seen in axisymmetric two-dimensional simulations. 

Earlier 2D, axisymmetric simulations (\citealt{scheck2008,marek_janka2009,hanke2012,summa2016}) have suggested that the SASI enhances neutrino energy deposition to promote explosion. However, comparisons with ray-by-ray and multi-dimensional neutrino transport (\citealt{skinner2016,dolence_2015,glas2018}) indicated that axisymmetric 2D simulations artifically enhance axial sloshing associated with the SASI to promote explosion. 

We emphasize that the development of a SASI $-$ in 3D as well as 2D $-$ is mainly restricted to failed explosions, with a smaller shock radius favorable to a faster growth rate via the advective-acoustic cycle (\citealt{foglizzo2002}). This is congruent with earlier work (\citealt{vartanyan2018b}) claiming that the SASI frequently appears in the context of delayed or failed explosions and more compact shock structures.

\begin{figure*}
    \centering
    \includegraphics[width=0.49\textwidth]{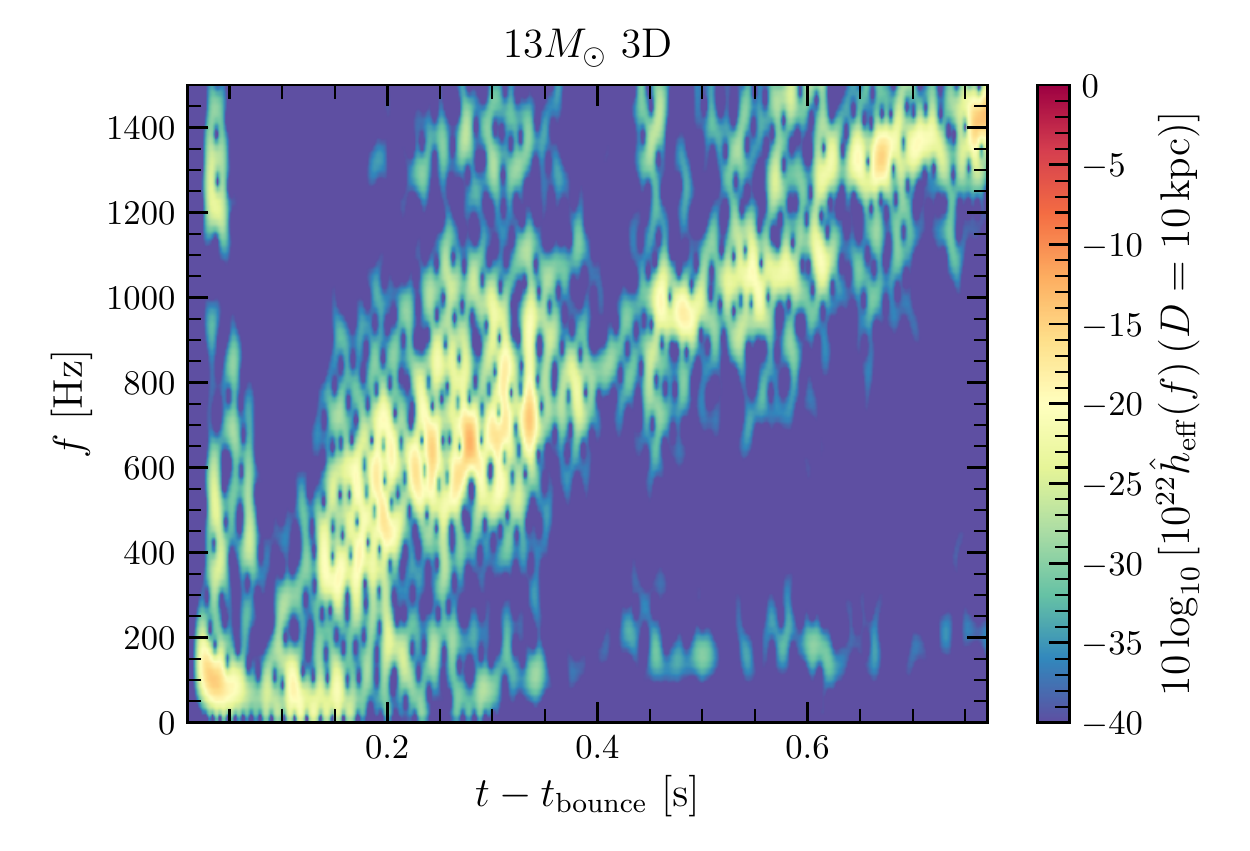}
    \includegraphics[width=0.49\textwidth]{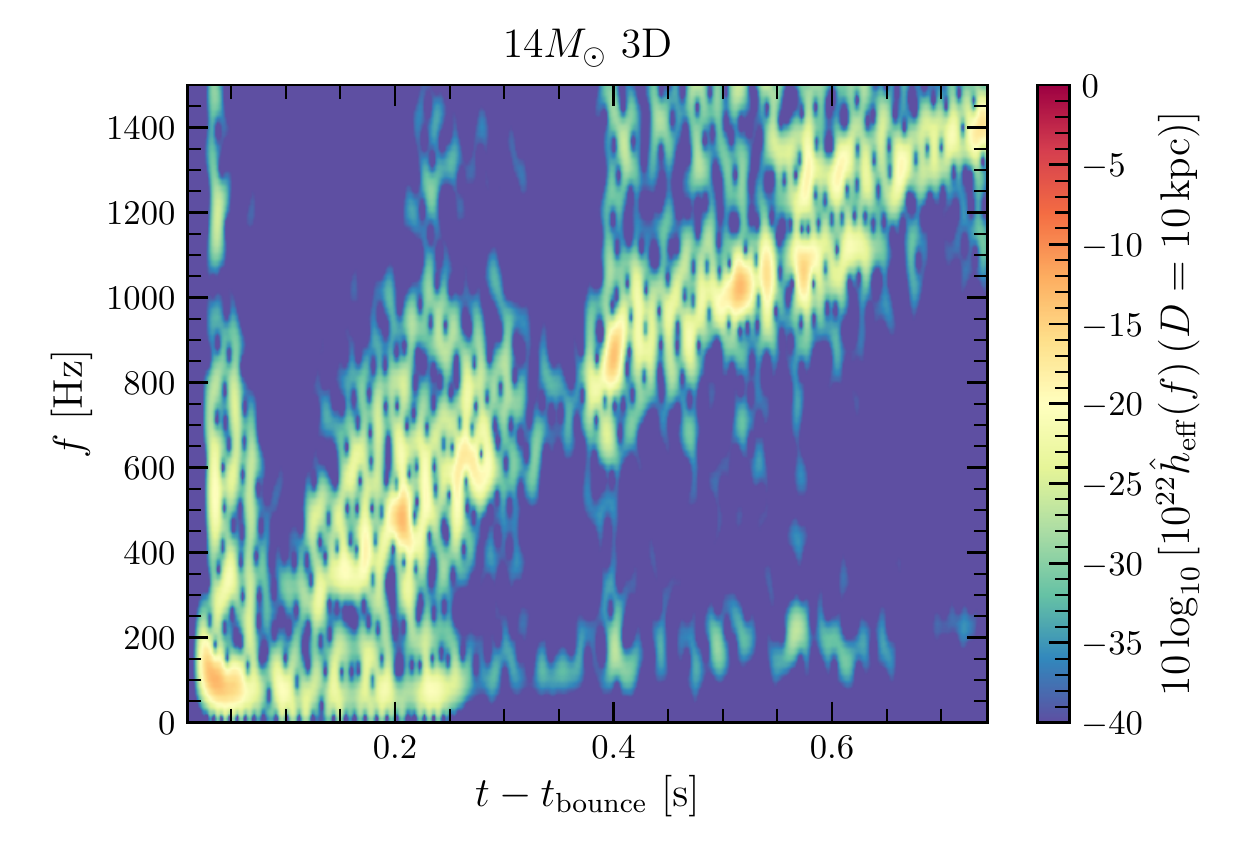}
    \hfill
    \includegraphics[width=0.49\textwidth]{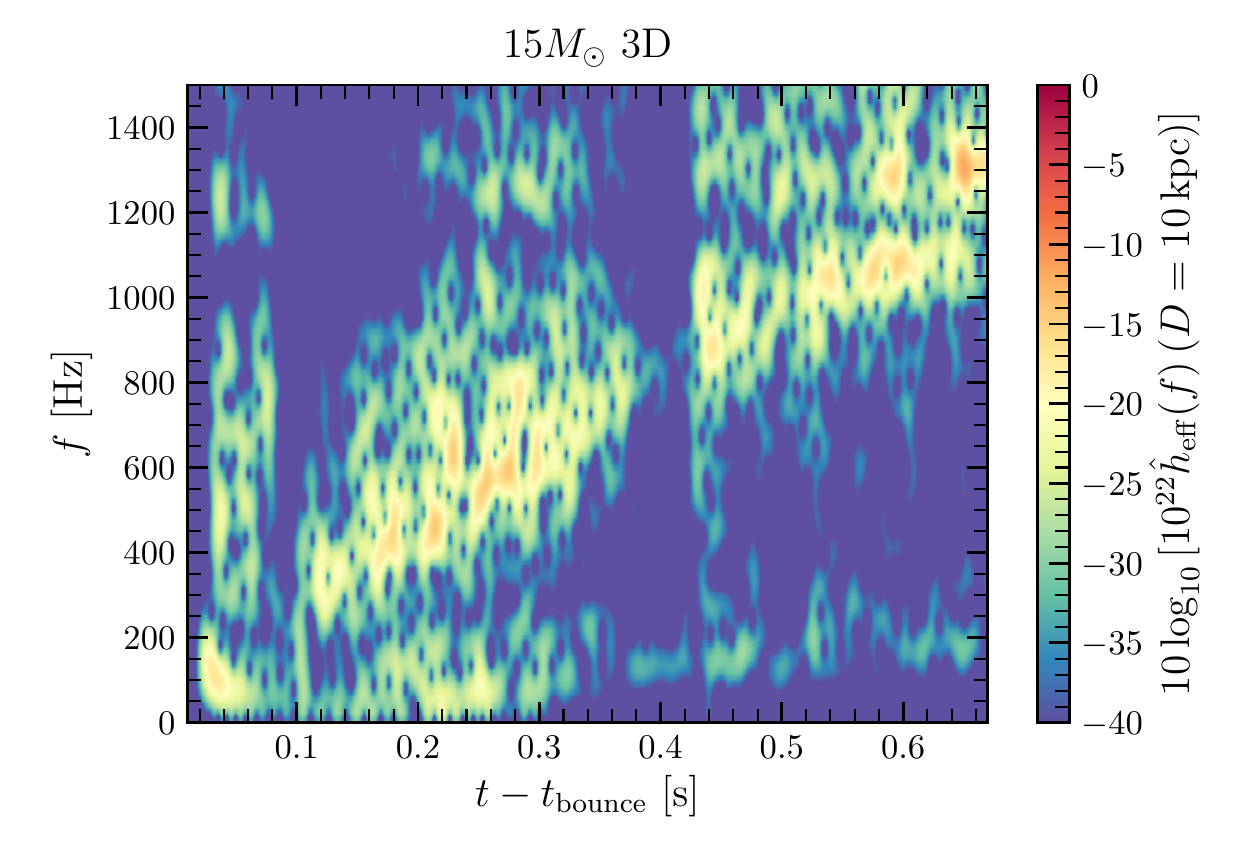}
    \includegraphics[width=0.49\textwidth]{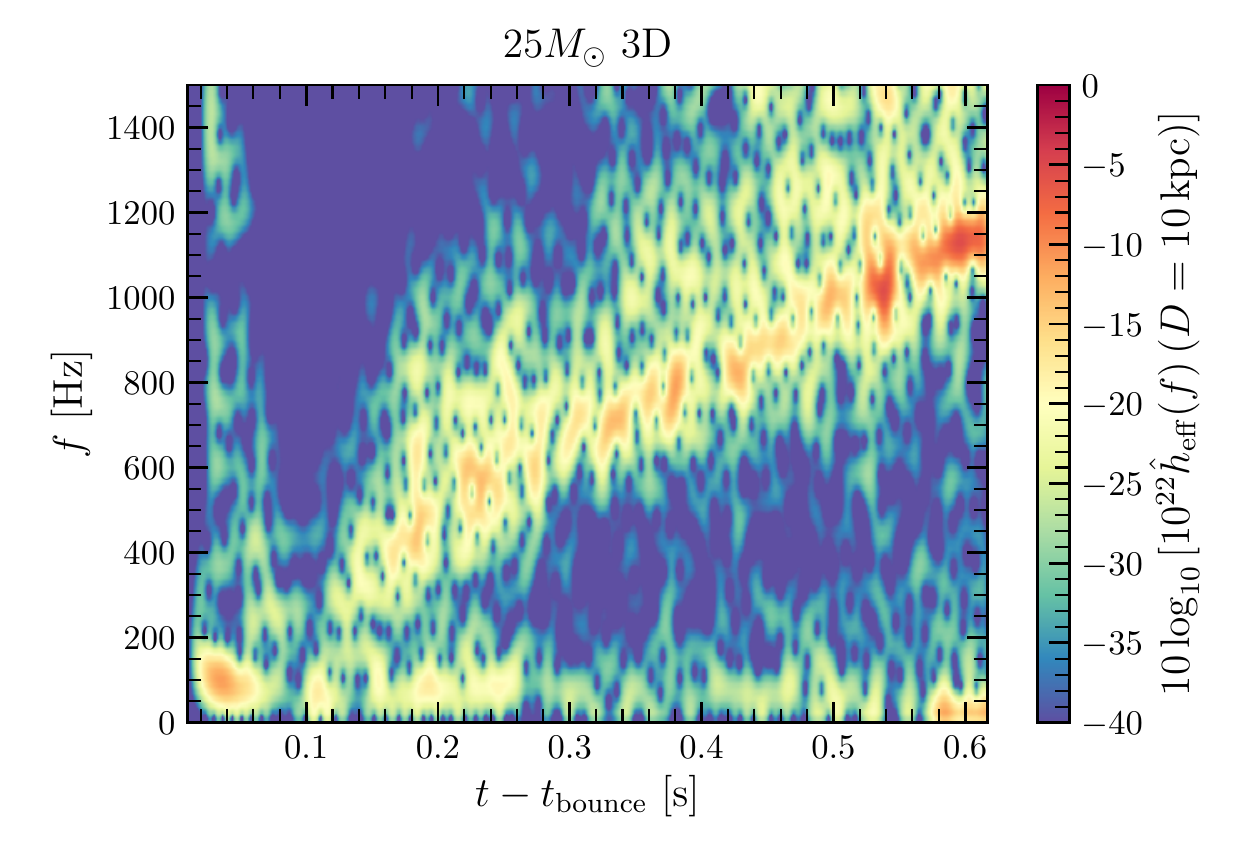}
    \caption{Gravitational wave spectrograms of the four models (13-, 14-, 15-, and 25-M$_{\odot}$) that exhibit some form of the SASI. Prior to 100 ms, we see the development of a low-frequency (less than 100 Hz) component associated with prompt convection. The fundamental mode frequency increases quadratically with time to 1 kHz by $\sim$500 ms postbounce (see also \protect\citealt{vsg2018}). Up to $\sim$250 ms, we see the telltale $\sim$100 Hz gravitational wave signature for all four models indicating the development of the SASI. These models either fail to explode, or explode late. After $\sim$400 ms, we see the development of a higher-frequency spiral SASI $-$ indicated by a gravitational wave signature at less than 200 Hz $-$ in the 13-, 14, and 15-M$_{\odot}$ progenitors, all of which fail to explode. The low-energy, low-frequency component after $\sim$300 ms in the 25-M${_\odot}$ progenitor does not correspond to the SASI, but is rather the linear memory due to an asymmetric explosion, and is visible in the spectrograms of all exploding models. The 25-M$_{\odot}$ progenitor explodes and shows no spiral SASI. The SASI signal is weaker than the $f$-mode frequency for all models considered. We see sub-millisecond power due to $f$- and $p$-modes in turbulent regions (visible in the spectrogram as power at frequencies greater than 1 kHz).}
    \label{fig:SASI_spec}
\end{figure*}

In Fig.\,\ref{fig:SASI_spec}, we portray gravitational-wave spectrograms of the four models (13-, 14-, 15-, and 25-M$_{\odot}$) that exhibit some form of the SASI. Prior to 100 ms, we see the development of a low-frequency (less than 100 Hz) component associated with prompt convection. The fundamental mode ($f$-mode) frequency increases with time to 1 kHz by $\sim$500 ms postbounce (\citealt{vsg2018}). Up to $\sim$250 ms after bounce, we see the telltale $\sim$100 Hz gravitational wave signature for these four models indicating the development of the the SASI. These models either fail to explode, or explode late. After $\sim$400 ms, we see the development of a spiral SASI $-$ indicated by a gravitational wave signature of less than $\sim$200 Hz $-$ in the 13-, 14-, and 15-M$_{\odot}$ progenitors, all of which fail to explode. The low-energy, low-frequency component after $\sim$300 ms in the 25-M${_\odot}$ progenitor does not correspond to the SASI, but is due to the long-term global motions, such as expansion mass asymmetries. This signature is present only in the exploding models. The 25-M$_{\odot}$ progenitor explodes and shows no spiral SASI. The SASI signal is weaker than the fundamental mode frequency for all models considered.

\begin{figure*}
    \centering
    \includegraphics[width=0.49\textwidth]{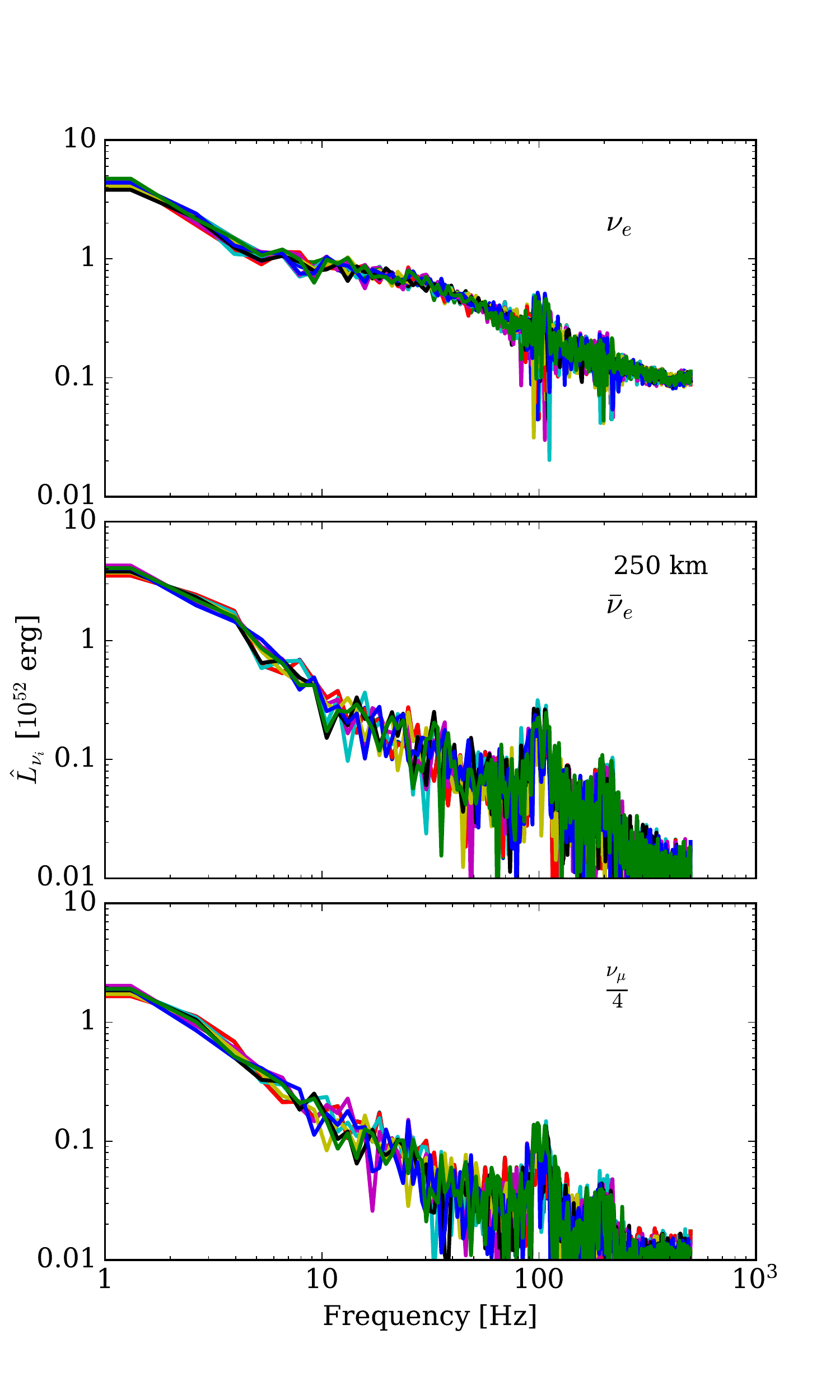}
       \includegraphics[width=0.49\textwidth]{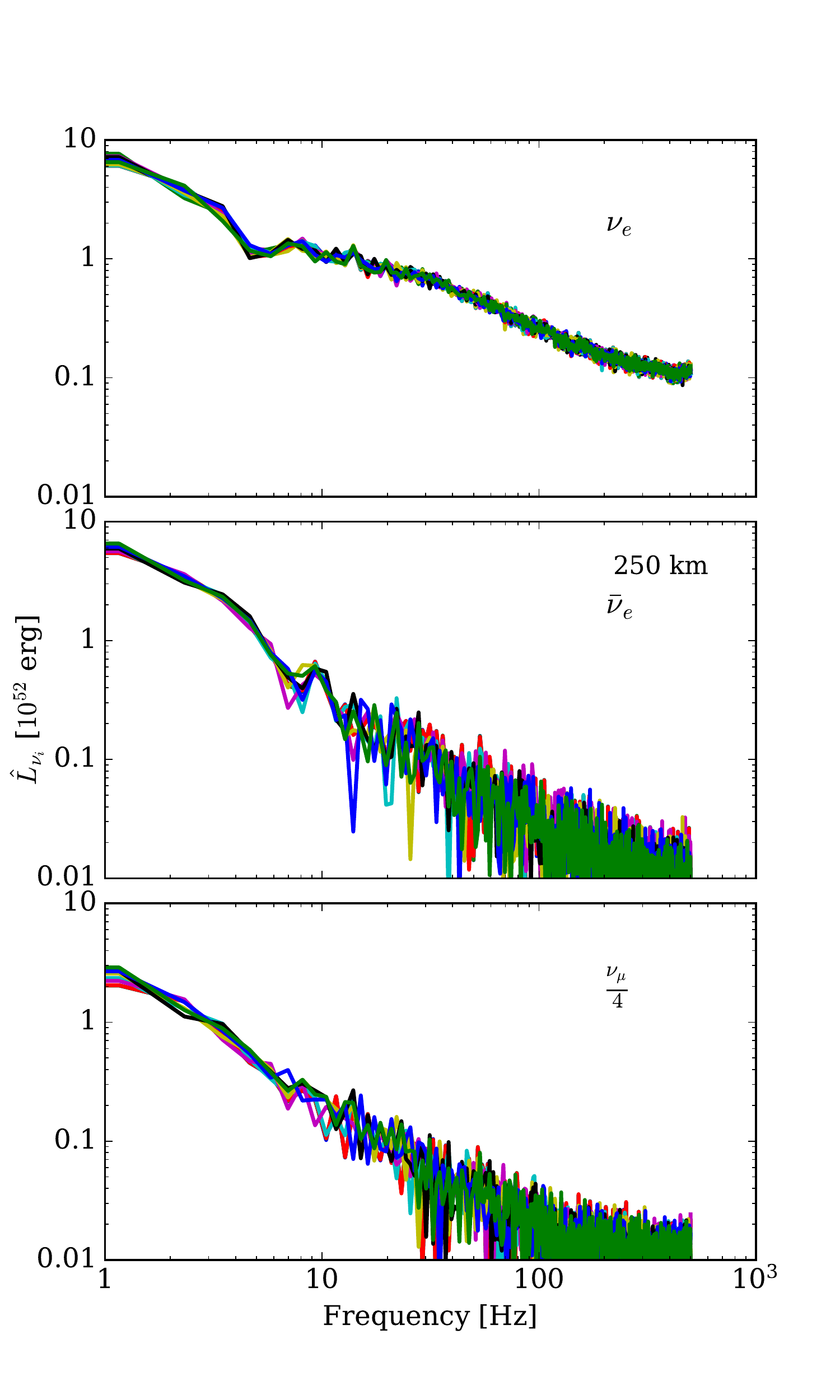}
\caption{We plot the Fourier transform $\hat{L}$ over an entire simulation of the luminosity at 250 km for the 13- (\textbf{left}, failed explosion) and 19-M$_{\odot}$ (\textbf{right}, successful explosion) progenitors along multiple lines-of-sight, indicated by the different colors, for all neutrino species. Note the strong peak in all species for the non-exploding 13-M$_{\odot}$ progenitor at $\sim$100 Hz (and a smaller peak at 200 Hz), indicative of the SASI. In the exploding 19-M$_{\odot}$ (\textbf{right}), we see a peak at 10 Hz, but no peak near $\sim$100 Hz. See text for a discussion.}
\label{fig:lum_FT}
\end{figure*}

In Fig.\,\ref{fig:lum_FT} we plot the Fourier transform of the luminosity for the 13- (failed explosion) and 19-M$_{\odot}$ (successful explosion) progenitors along multiple, arbitrarily chosen, lines-of-sight, indicated by the different colors, for all neutrino species. Note the strong peak at $\sim$100 Hz in all  neutrino species for the non-exploding 13-M$_{\odot}$ progenitor at $\sim$100 Hz, indicative of the SASI. In the exploding 19-M$_{\odot}$ (right panel), we see a clear peak at $\sim$10 Hz, indicative of large-scale explosion asymmetries, and the $\sim$100 Hz SASI signal is absent.

\cite{summa2018} find dynamic shock expansion due to kinetic energy deposition in the SASI spiral arm, driving their models to explosion. In our (non-rotating models), we do not witness shock revival for the three failed models where the spiral SASI does develop. Therefore, when the SASI appears in our non-rotating models, it is usually in the context of receding shocks and failed explosions. The turbulence seen is always predominantly a consequence of neutrino-driven convection, and exploding models rarely show any signs of the SASI, at least for our non-rotating model set.

\section{LESA}\label{sec:LESA}

The lepton-number emission self-sustained asymmetry (LESA) was proposed by \cite{2014ApJ...792...96T} as a neutrino-hydrodynamical instability resulting in $\nu_e - \bar{\nu}_{e}$ emission asymmetry, with possible implications for nucleosynthesis (\citealt{fujimoto2019}). In later work, \cite{vartanyan2018a,oconnor_couch2018b} identified the LESA by examining the dipole component of the spherical harmonic decomposition of the net lepton number flux (F$_{\nu_e}$ $-$ F$_{\bar{\nu}_{e}}$) for a single simulation in 3D of the 16-M$_{\odot}$ progenitor (\citealt{wh07}). 
\cite{walk2018} found that, for rotating models, the LESA instability is suppressed associated with 
weaker PNS convection and \cite{walk2018}, \cite{glas2018}, and \cite{walk2019} have studied possible connections 
between neutrino emissions, neutrino oscillations, and the LESA.

\begin{figure*}
    \centering
    \includegraphics[width=0.7\textwidth]{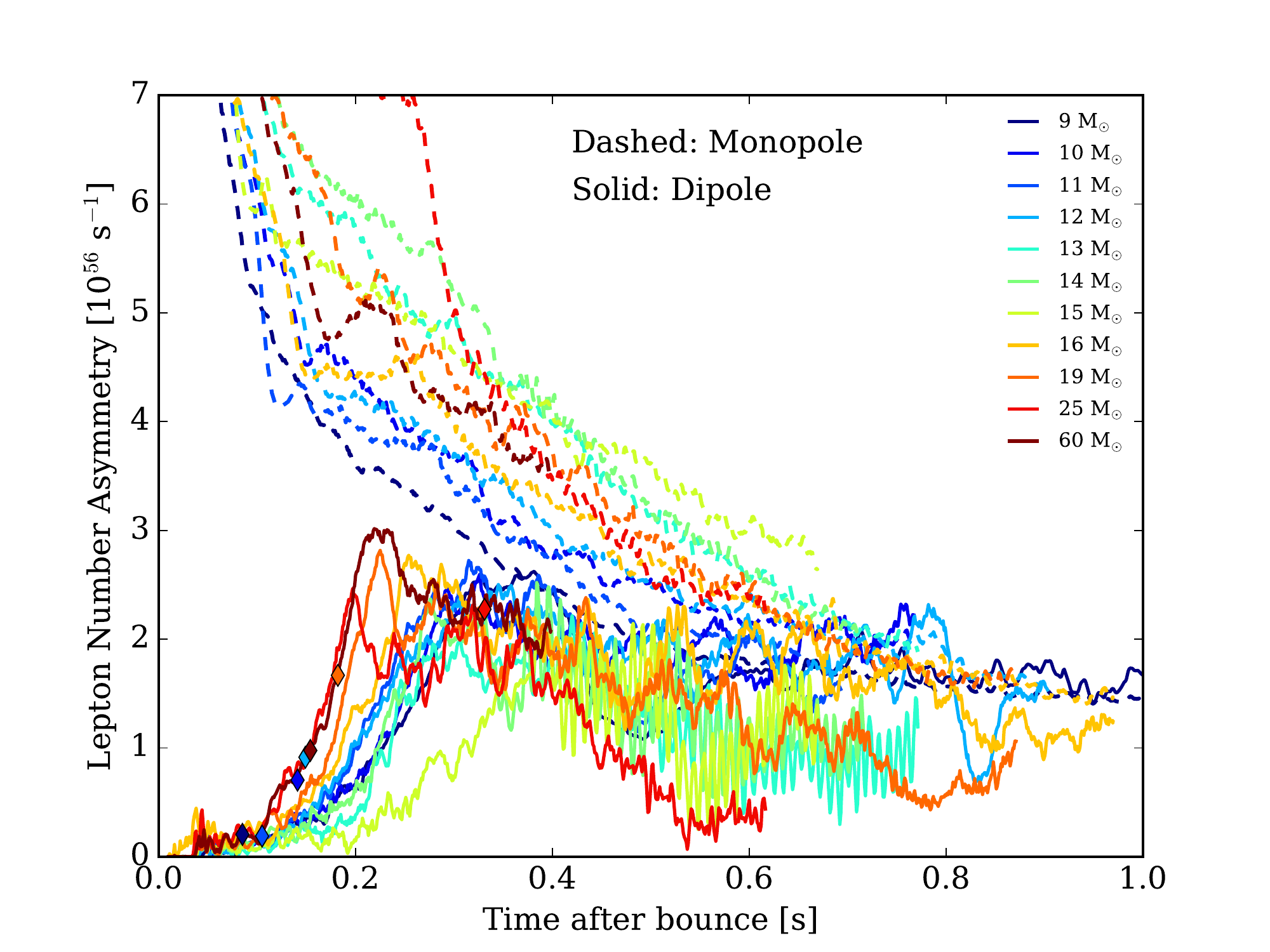}
    \caption{The monopole and dipole components of the LESA (at 500 km) as a function of time after bounce (in seconds). In all models, we see the development of the LESA at or shortly after $\sim$200 ms. The LESA disappears after $\sim$500 ms for the 25-M$_{\odot}$ progenitor and abates in magnitude for all but the lowest mass progenitors.}
    \label{fig:LESA}
\end{figure*}

We now extend our exploration of the possible presence of the LESA to 11 progenitors evolved in 3D. In Fig.\,\ref{fig:LESA}, we depict the monopole and dipole components of the lepton asymmetry as a function of time after bounce at 500 km. Here, we follow \cite{oconnor_couch2018a} and plot the dipole magnitude,
\begin{equation}
A_{\mathrm{dipole}} = 3 \times \sqrt{\sum_{i=-1}^{1} a_{1i}^2},,
\end{equation} using the normalization scheme of \cite{burrows2012}.

For all of our models $-$ irrespective of explosion outcome $-$ we see the development of the LESA, and illustrate the monopole and dipole components of this asymmetry in Fig.\,\ref{fig:LESA}, consistent with recent 3D simulations (\citealt{oconnor_couch2018a}; \citealt{glas2018}). Note the strong periodicity in the non-exploding models (13-, 14-, 15-M$_{\odot}$) after 400 ms, as the spiral SASI develops. We do not find strong evidence that the LESA correlates with either the behavior of the shock surface or the accretion rate; rather, as we note in \S\,\ref{sec:corr}, it is the neutrino luminosity itself that correlates with the shock radius and accretion rate temporal oscillations, in agreement with \cite{dolence_2015}. 

\begin{figure*}
    \centering
    \includegraphics[width=0.85\textwidth]{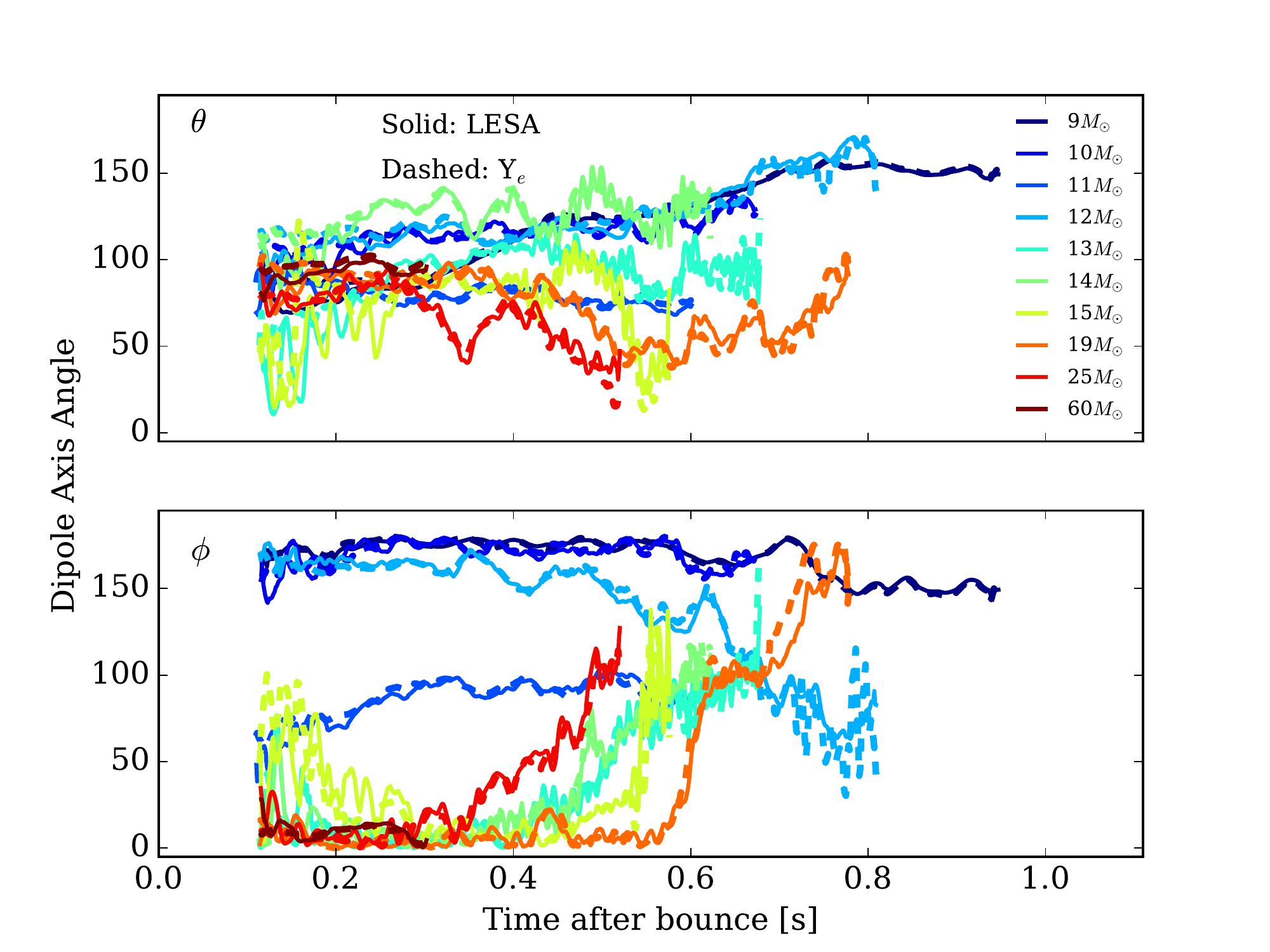}
     \caption{The orientation of the LESA dipole (at 500 km) and the Y$_\mathrm{e}$ dipole (measured at 25 km, tracing PNS convection) for the various models considered in this paper. For all models, we see that the LESA evolution closely correlates with the Y$_\mathrm{e}$ evolution. It trails the Y$_\mathrm{e}$ evolution by $\sim$2 ms, corresponding to the light travel time from the PNS to 500 km. See text for a discussion.}
         \label{fig:LESA_Ye_25}
\end{figure*}

In Fig.\,\ref{fig:LESA_Ye_25}, we plot the $\theta$ and $\phi$ components of the orientation of the LESA dipole axis, and of the radius-weighted dipole axis of the electron fraction distribution at 25 km, following the prescription of \cite{oconnor_couch2018b}. For all eleven models, we see remarkable correlation between the orientation of the LESA dipole axis and the dipole axis of the electron-fraction distribution in the convective PNS. We see that the Y$_\mathrm{e}$ dipole precedes that of the LESA by $\sim$2 ms, which can be explained by the light travel time from the PNS to 500 km, where we measure the LESA. These results build on recent evidence (\citealt{oconnor_couch2018b,glas2018}) to suggest that hemispheric differences in PNS convection drive the LESA dipole.

\section{Neutrino and Gravitational Wave Emission Correlations}\label{sec:corr}

\begin{figure}
    \centering
    \rotatebox[origin=c]{-90}{%
    \begin{minipage}{1.2\textwidth}
    \centering
        \includegraphics[width=1.05\linewidth]{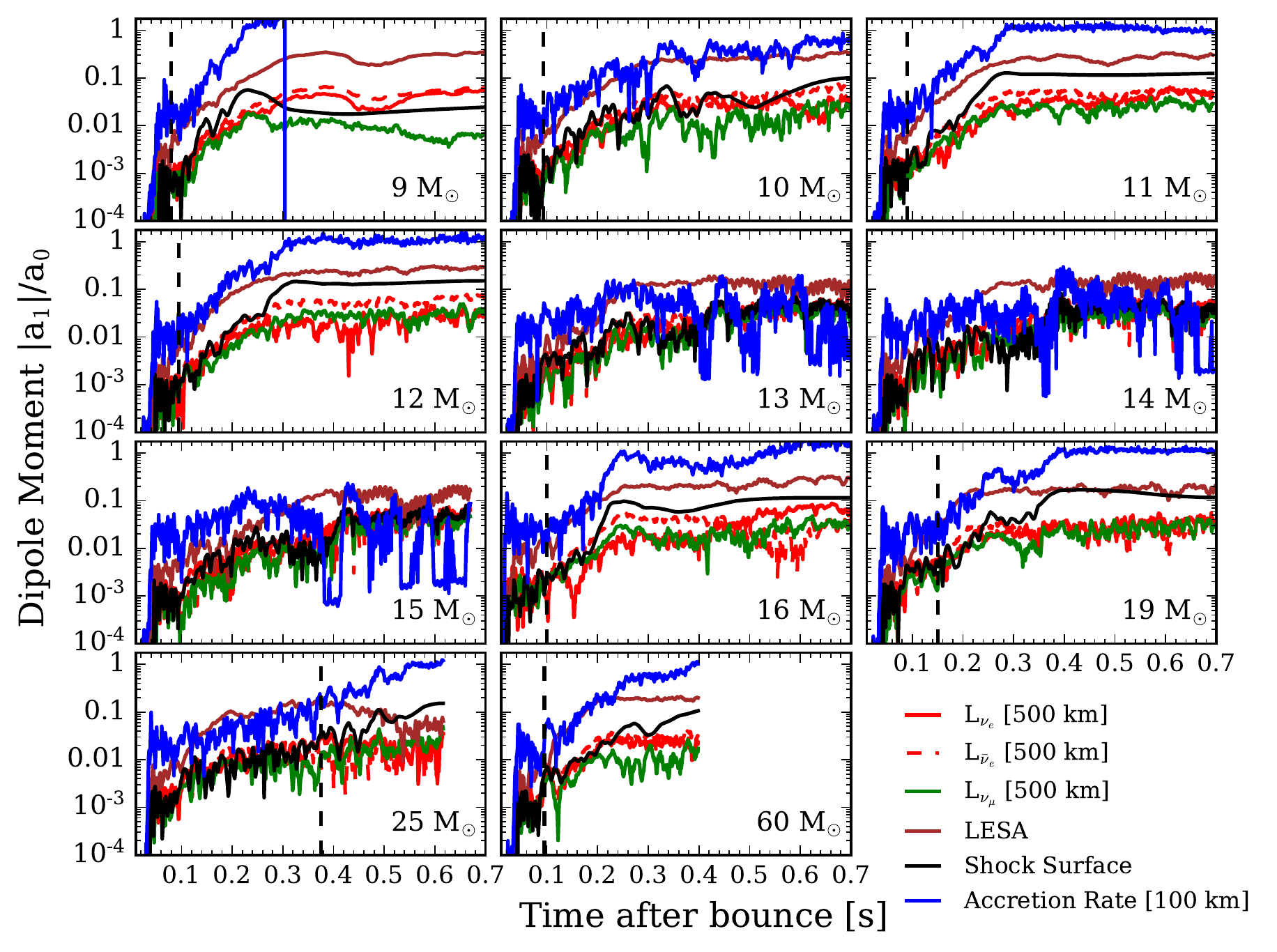}
    \caption{The normalized dipole components of the accretion rate (blue, at 100 km), shock surface (black), and neutrino luminosities (solid-red for electron-neutrinos, dashed-red for electron anti-neutrinos, and green for heavy-neutrinos, at 500 km). We see a hierarchy, with the normalized accretion rate dipole being largest, and that for the neutrino luminosity smallest. Among the neutrino luminosities, the electron-neutrino and anti-neutrino luminosities have comparable normalized dipoles, with the anti-neutrino dipole slightly larger, and the heavy-neutrino having the smallest dipole. This may attest to the different positions of the neutrinospheres for the different neutrino species. Notably, the non-exploding models have smaller accretion rate dipoles.}
    \label{fig:lum_rad_mdot}
    \end{minipage}
    }
\end{figure}

In this section, we explore correlations of the potentially observable neutrino and gravitational wave signatures with the inner dynamics of the supernova. In Fig.\,\ref{fig:lum_rad_mdot}, we plot the normalized (by the monopole) dipole components of the accretion rate (blue, at 100 km), shock surface (black), LESA (brown),  and neutrino luminosities (solid-red for electron-neutrinos, dashed-red for electron anti-neutrinos, and green for heavy-neutrinos, at 500 km). 
 We use the approach outlined in \cite{burrows2012} to
decompose the shock surface R$_s(\theta,\phi)$ into spherical harmonic components with coefficients:
\begin{equation}\label{eq:alm}
a_{lm} = \frac{(-1)^{|m|}}{\sqrt{4\pi(2l+1)}} \oint R_s(\theta,\phi) Y_l^m(\theta,\phi) d\Omega\, ,
\end{equation} normalized such that $a_{00} = a_{0} =\langle R_s\rangle$ (the average
shock radius). $a_{11}$, $a_{1{-1}}$, and $a_{10}$ correspond to the average
Cartesian coordinates of the shock surface dipole $\langle x_s\rangle$, $\langle y_s\rangle$,
and $\langle z_s\rangle$, respectively.
The orthonormal harmonic basis functions are given by
\begin{equation}
Y_l^m(\theta,\phi) = \begin{cases}
        \sqrt{2} N_l^m P_l^m(\cos\theta) \cos m\phi&            m>0\, ,\\
        N_l^0 P_l^0(\cos\theta) &                               m=0\, ,\\
        \sqrt{2} N_l^{|m|} P_l^{|m|}(\cos\theta) \sin |m|\phi&  m<0\, ,
\end{cases}
\end{equation}
where
\begin{equation}
N_l^m = \sqrt{\frac{2l+1}{4\pi}\frac{(l-m)!}{(l+m)!}}\, ,
\end{equation} $P_l^m(\cos\theta)$ are the associated Legendre polynomials,
and $\theta$ and $\phi$ are the spherical coordinate angles. We define the norm,

\begin{equation}
A_{\ell} = \frac{\sqrt{\sum_{m=-\ell}^{\ell} a_{\ell m}^2}}{a_{00}}\,.
\end{equation}

We see a hierarchy, with the normalized accretion rate dipole being largest, and that for the neutrino luminosity smallest. Non-exploding models have smaller accretion rate dipoles. We see that $\nu_e$ and $\bar{\nu}_e$ neutrino luminosities have comparable normalized dipoles, with the $\bar{\nu}_e$ dipole slightly larger, and the $\nu_\mu$ neutrino luminosity having the smallest dipole. This may attest to the different neutrinosphere radii for the different neutrino species, though the temporal variations of their luminosities track each other. We see remarkable similarity between the oscillations in the luminosity dipole and shock surface dipole, with the former lagging by $\sim$5$-$10 ms at early times, prior to explosion, due to advection of accreta from the stalled shock to the neutrinosphere. The shock radius dipole has longer period variations, of $\sim$10 ms and greater, and settles earliest after explosion, asymptoting to a roughly constant value in those models whose explosion energies have also begun to asymptote (\citealt{radice2019}). 
The shock surface lies interior to $\sim$500 km at the early times plotted here, out to $\sim$300 ms. The 25-M$_{\odot}$ progenitor shows the clearest correlation between accretion and luminosity; its later explosion is powered by a higher sustained accretion rate and consequent neutrino luminosity. 

To correlate the time variation of the shock radii, the accretion rate, and the neutrino luminosities, we investigated their Fourier frequency content. Prior to explosion, we generally see large amplitude variation for these physical quantities on short timescales of $\sim$5 ms. After explosion, the dipole of the shock radii, neutrino luminosities, and accretion rates transition to small amplitude, long-period variations. In the accretion rate and neutrino luminosities, we see fast temporal variation with timescales of $\sim$5 ms, within a broader, quasi-periodic envelope with a typical width of $\sim$40 ms but as high as $\sim$100 ms, corresponding to large-scale anisotropies of the shock motion modulating the accretion rate. The 9-M$_{\odot}$ progenitor, whose explosion proceeds relatively isotropically and whose accretion phase ends early, lacks such a feature. Additionally, for the first $\sim$400 ms, of $-$ for example, $-$ the 25-M$_{\odot}$ progenitor's evolution, we see variation on $\sim$30 ms timescales, corresponding to advection of material from the (initially) slowing growing shock. 

\begin{figure*}
    \centering
    \hfill
    \includegraphics{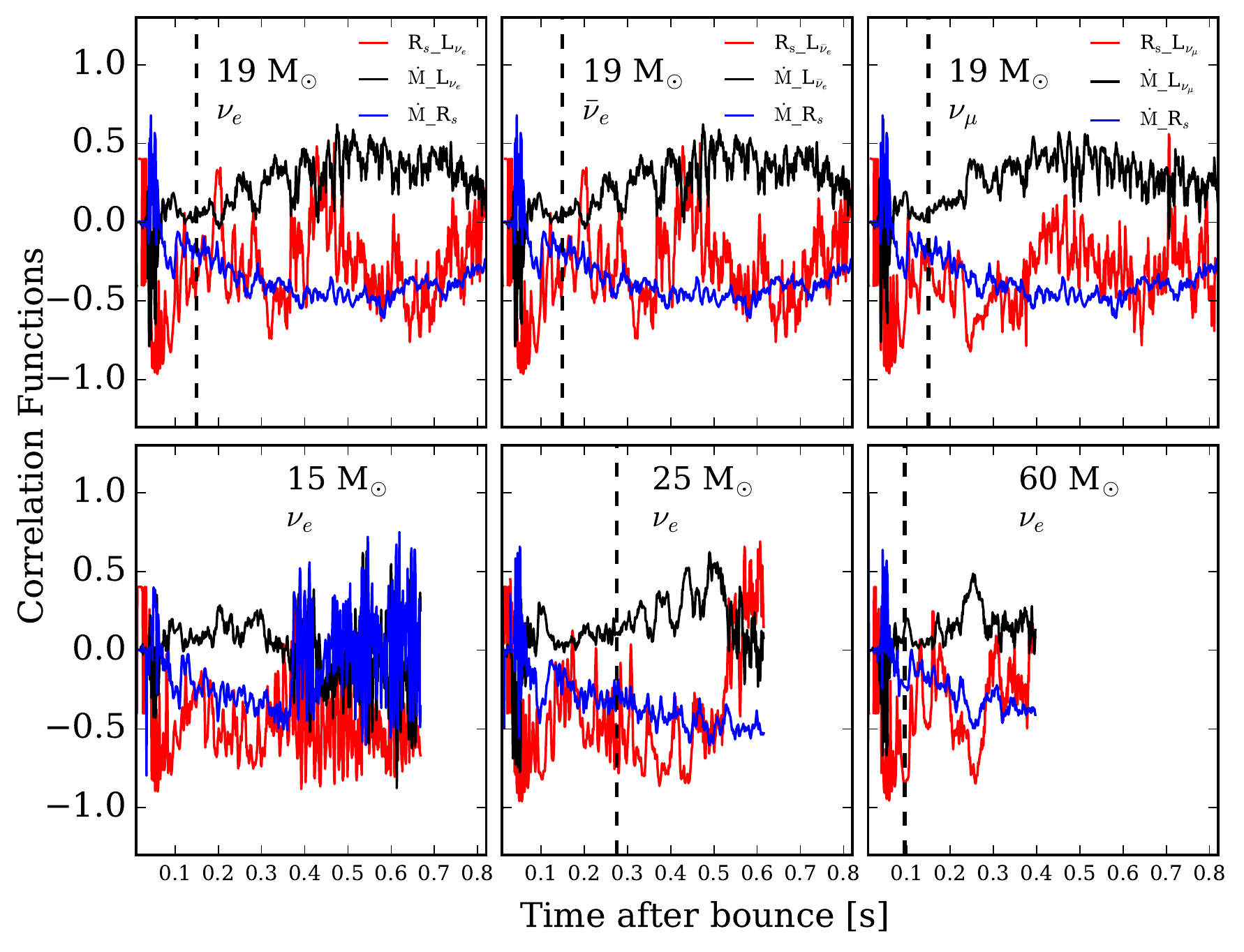}
    \caption{Angle-correlations between the various physical quantities depicted for the 19-M$_{\odot}$ progenitor (\textbf{top}) and other various progenitors (\textbf{bottom}). From left to right in the top panel, these correlations are shown for different neutrino species for the 19-M$_{\odot}$. We note the strong dependence of heavy-neutrinos on the accretion rate. We see correlation between the luminosity and accretion rate, and anti-correlation between the luminosity and shock radius, and shock radius and accretion rate for all models, exploding and non-exploding alike. After $\sim$300 ms, the non-exploding models (the 15-M$_{\odot}$ progenitor is shown here) show no persistent correlation, but rather high-frequency oscillation around zero, indicative of SASI activity. See text for a discussion.}
    \label{fig:angle_corr}
\end{figure*}

For the non-exploding models, the shock radius, accretion rate, and neutrino luminosity all sustain persistent, short-timescale variations even at late times, $\sim$500 ms postbounce. Furthermore, we see in non-exploding models a transition to shorter timescales and higher frequencies (see Fig.\,\ref{fig:angle_corr} and following) after $\sim$400 ms, coincident with the development of the spiral SASI thereabouts. This is most visible in the 14-M$_{\odot}$ model, for which we see periodic $\sim$10-ms variations after $\sim$400ms in the shock surface, accretion rate, and neutrino luminosity dipoles (for all species). After the spiral SASI develops, we also see large drops by two orders of magnitude in the dipole of the accretion rate over $\sim$80 ms timescales. Concurrently, and for the non-exploding models alone, we see the development of periodic $10$-ms oscillations of the LESA dipole, which we associate with the spiral motion of the SASI, perhaps due to modulation of infalling accretion. The LESA dipole shows little to no temporal variation for the exploding models. While the 25-M$_{\odot}$ progenitor does show early SASI activity, it explodes without a spiral SASI developing. We reiterate that $\sim$10 ms periodicity of the neutrino luminosity several hundred milliseconds after bounce is an indicator of spiral SASI activity.

We follow the formalism of \cite{kuroda2017} to calculate the time-dependent angle-integrated correlation $X\,(t)$ between two physical quantities, A$_1\,(\Omega)$ and A$_2\,(\Omega)$,

\begin{equation}
X\,(t) =  \frac{\int A_1\,(t,\Omega)\, A_2\,(t,\Omega)\, d\Omega}{\sqrt{\int A_1\,(t,\Omega)^2 d\Omega  \int A_2\,(t,\Omega)^2\,d\Omega}}\,.
\end{equation} In Fig.\,\ref{fig:angle_corr}, we plot angle-averaged correlations as a function of time after bounce between the shock radius and neutrino luminosity, shock radius and accretion rate, and accretion rate and neutrino luminosity for various progenitors.  In the top panel, we show the correlation with the neutrino luminosities of different neutrino species for the 19-M$_{\odot}$ progenitor. We note the dependence of the heavy-neutrino luminosity on the accretion rate. We see correlation between the luminosity and accretion rate, and shock radius and accretion rate, and weaker anti-correlation both between the luminosity and shock radius, for all models exploding and non-exploding alike. However, after $\sim$300 ms, the non-exploding models show no persistent correlation, but rather show variation around zero on $\sim$10-ms timescales, illustrating SASI activity.

\begin{figure*}
    \centering
         \includegraphics[width=0.85\textwidth]{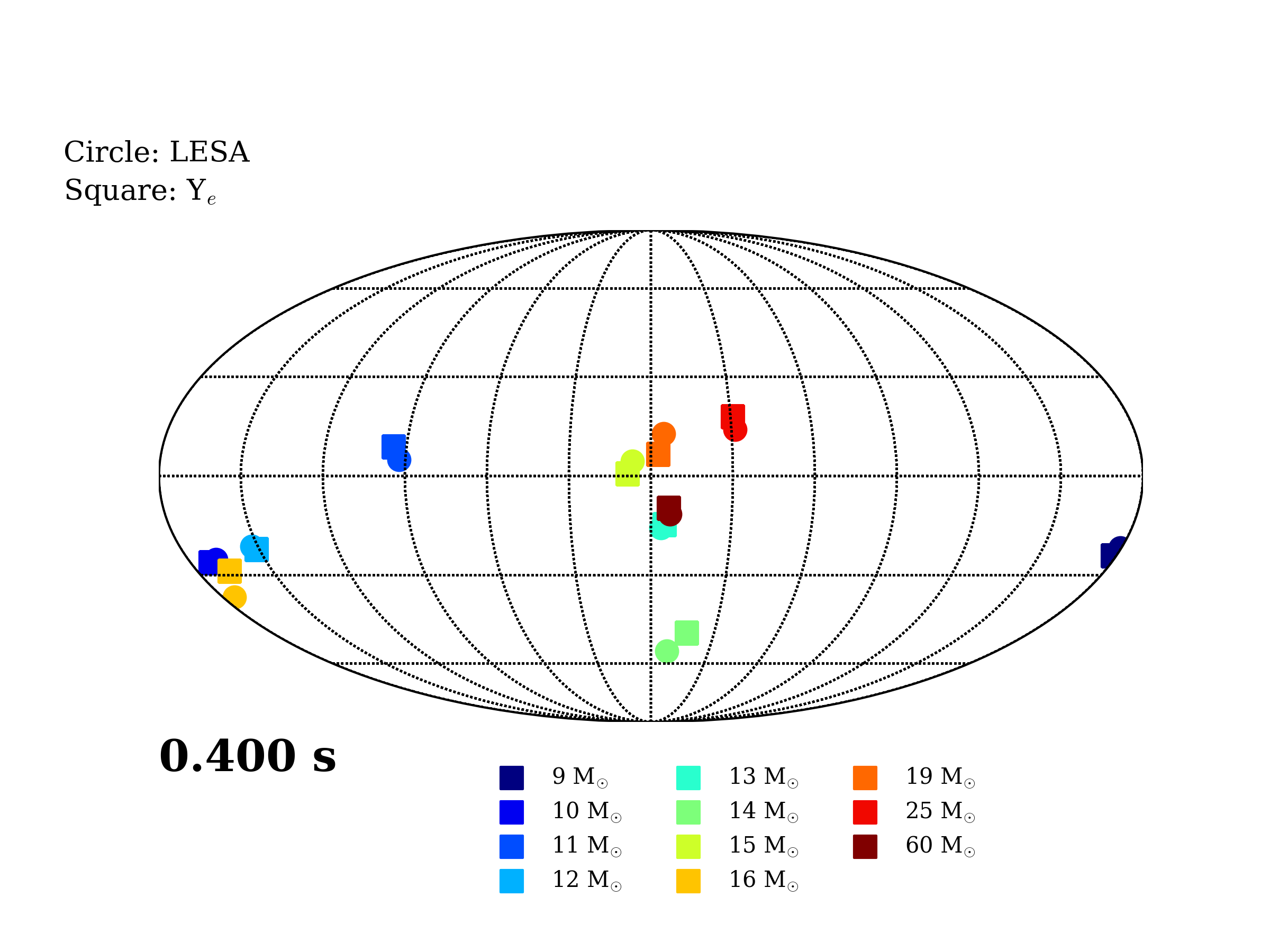}
        \includegraphics[width=0.85\textwidth]{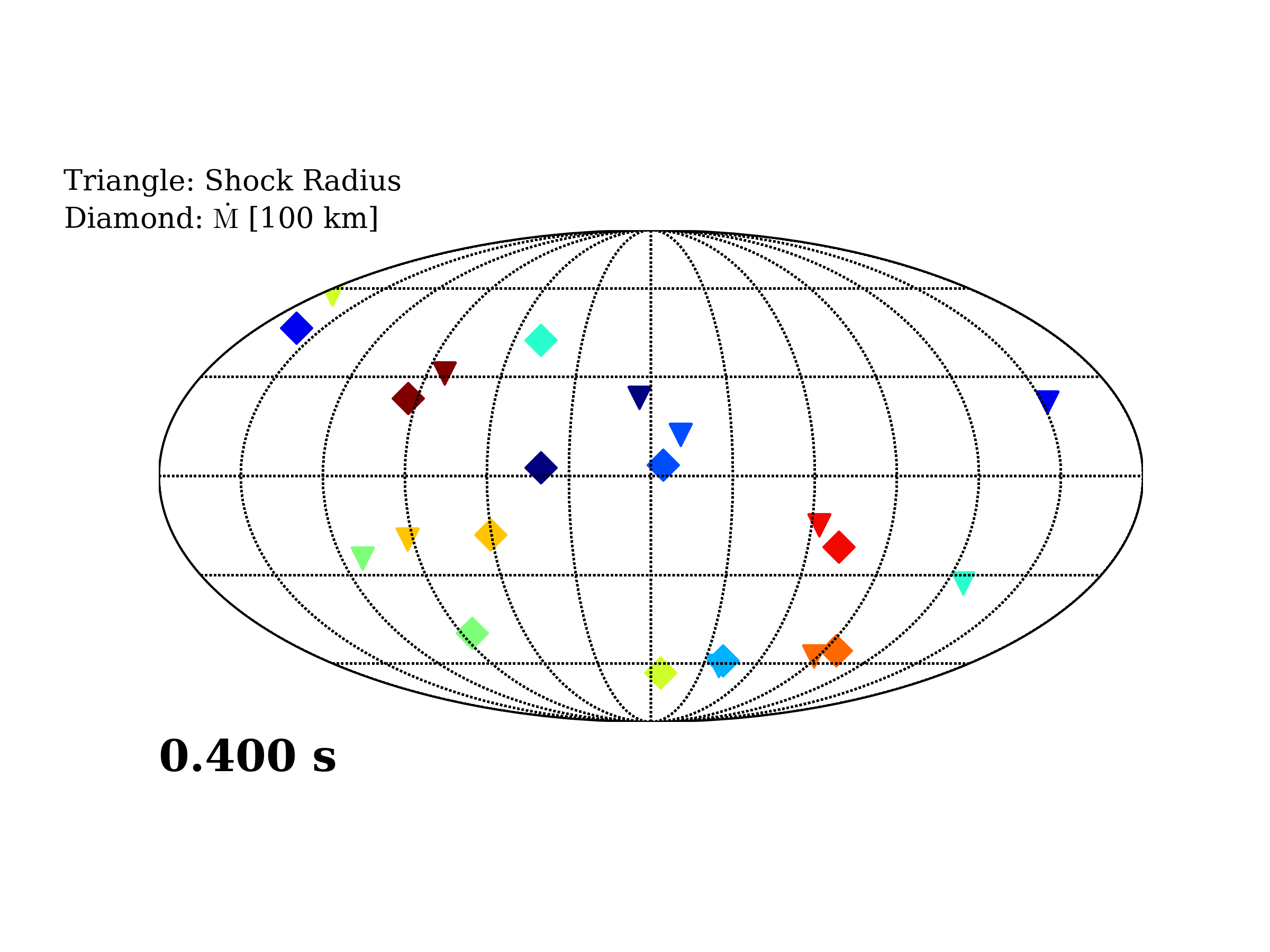}
    \caption{\textbf{Top}: Mollweide projection of the dipole
    directions of the LESA at 500 km (square) and Y$_\mathrm{e}$ (circle) in the convective PNS at 25 km at 400 ms postbounce. \textbf{Bottom}: Same as above, but for the shock radius (triangle) dipole and the accretion rate anti-dipole at 100 km (diamond) 400 ms after bounce for all progenitor models studied in this paper. The LESA dipole closely traces the Y$_\mathrm{e}$ dipole when the LESA is active. Similarly, for the exploding models, the shock radius dipole strongly anti-correlates with the accretion rate.}
        \label{fig:moll}
\end{figure*}

In Fig.\,\ref{fig:moll}, we visualize the above angle correlations by plotting a Mollweide projection of the dipole directions of the LESA at 500 km (square), Y$_\mathrm{e}$ (circle) in the convective PNS at 25 km, and the shock radius (triangle), as well as of the accretion rate anti-dipole direction at 100 km (diamond) at a snapshot 536 milliseconds after bounce for all progenitor models studied in this paper. The LESA dipole closely traces the Y$_\mathrm{e}$ dipole when the LESA is active. For the non-exploding models, we see precession of the LESA dipole direction around the Y$_\mathrm{e}$ dipole direction as the spiral SASI develops. In the bottom panel, we plot the antipode of the accretion rate to illustrate its strong anti-correlation in direction with the shock radius.

\begin{figure*}
    \centering
    \includegraphics[width=0.48\textwidth]{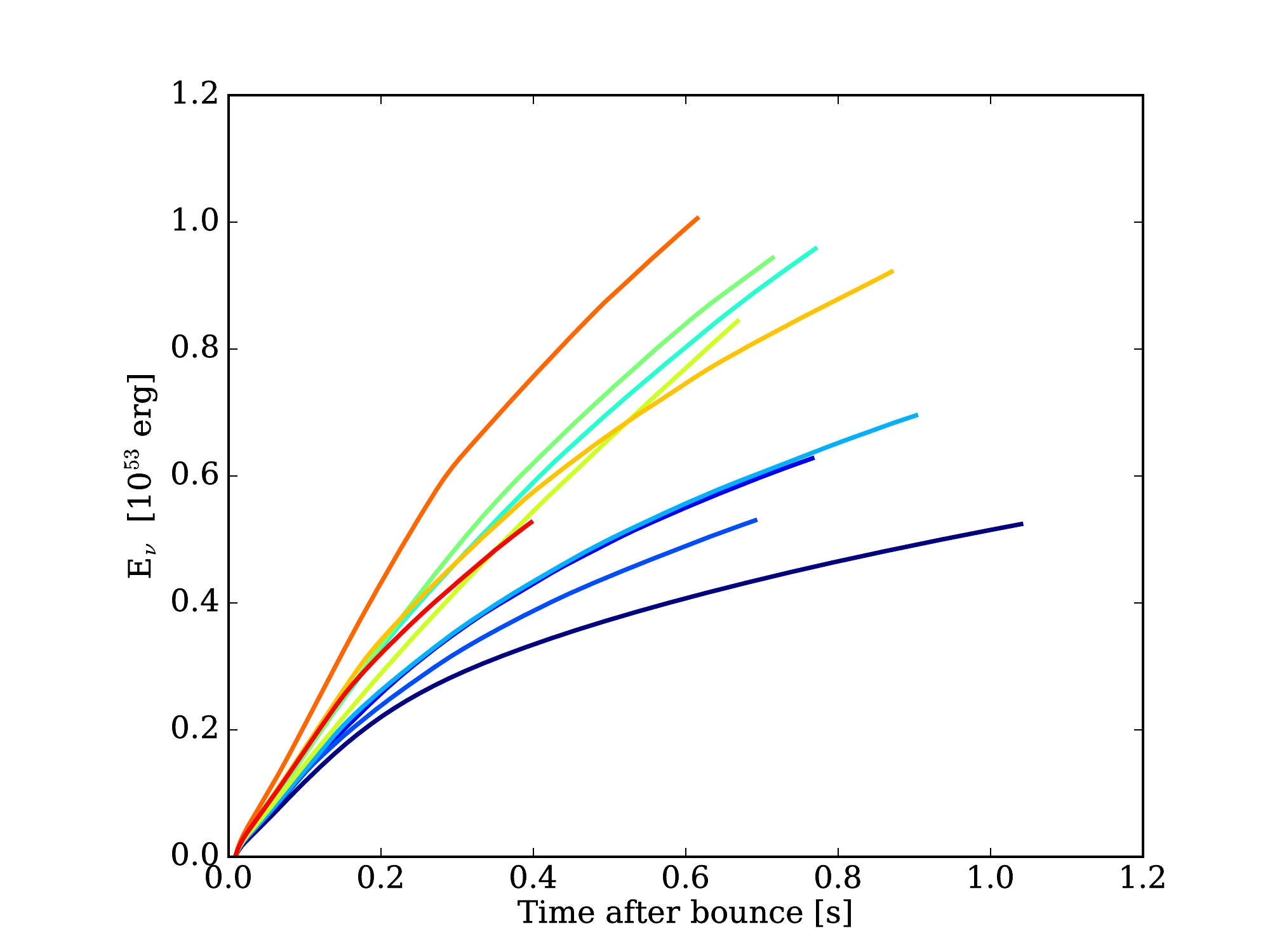}
    \includegraphics[width=0.48\textwidth]{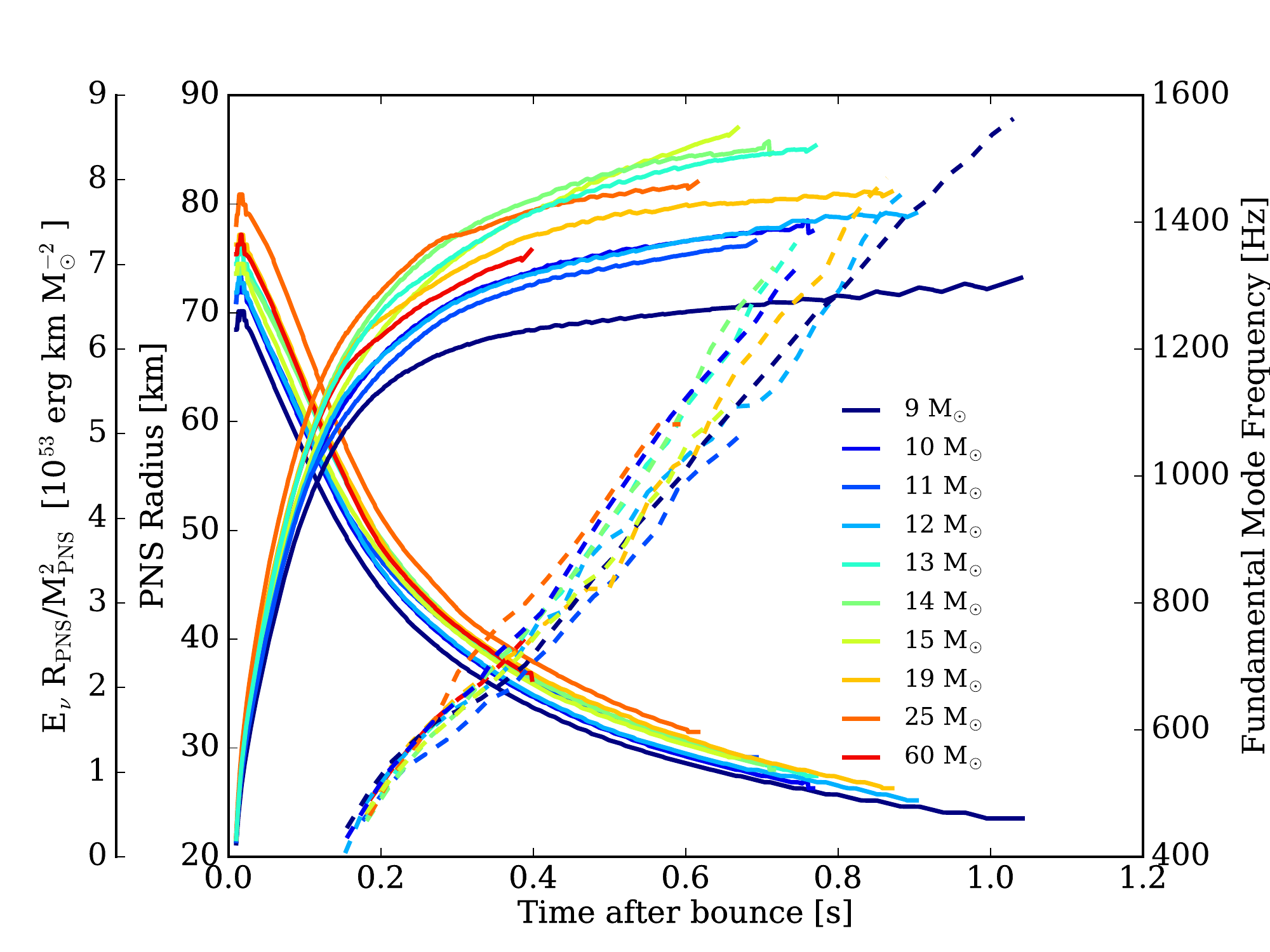}
    \caption{\textbf{Left}: The total neutrino energy loss as a function of time after bounce (in seconds) for all progenitors highlighted in this paper. \textbf{Right}: The PNS radius (km) on the left y-axis and the gravitational wave fundamental mode frequency (dashed lines), right y-axis, as a function of time after bounce (in seconds). We also plot the total neutrino energy loss normalized by the Newtonian binding energy of the PNS. As the PNS radiates neutrinos and loses energy, it contracts and the fundamental ($f$-) mode frequency increases. Neglecting the 9-M$_{\odot}$ progenitor, normalizing energy-loss by PNS mass results in a smaller variation, $\sim$10\%, by progenitor mass. Roughly, we see a correlation between highest neutrino energy loss (\textbf{left}), and smallest PNS radii (\textbf{right}). As seen in \protect\cite{vsg2018}, at later times the fundamental mode frequency begins to turn over quadratically (most visible here for the 9-M$_{\odot}$, carried out furthest). However, models need to be carried out to later postbounce times for this to be easily discernible.} 
    \label{fig:lum_pns_gw}
\end{figure*}

In Fig.\,\ref{fig:lum_pns_gw}, we plot the PNS radius, integrated energy lost by neutrinos, and the fundamental mode frequency of the emitted gravitational waves as a function of time. Increasing neutrino energy losses cause the core to shrink and the fundamental mode frequency to increase. Both the fundamental mode frequency and the PNS radii show less than $\sim$10\% scatter with progenitor mass. The total energy lost is normalized by the Newtonian binding energy of the PNS to reduce scatter. Four hundred milliseconds after bounce, the PNS's have typical radii of $\sim$38 km and $f$-mode gravitational wave frequencies of $\sim$800 Hz. To illustrate the correlation between the PNS radii and $f$-mode frequency, we find a least-squared fit to the mean behavior,

\begin{equation}
    R_{\mathrm{PNS}} [\mathrm{km}] \approx 46 \times \left(\frac{f [\mathrm{kHz}]}{1.3} - 0.23\right)^{-0.25}\,,
\end{equation} where the best-fit power lies between 0.25 and 0.31, or conversely, we can invert to find that the fundamental mode frequency scales as the PNS radius to the 3 to 4 power, to a multiplicative constant and additive offset. We emphasize that this scaling is for illustrative purposes, to indicate the PNS-probing power of future supernovae GW detections, and can certainly be improved upon.

\begin{figure*}
    \centering
    \includegraphics[width=0.55\textwidth]{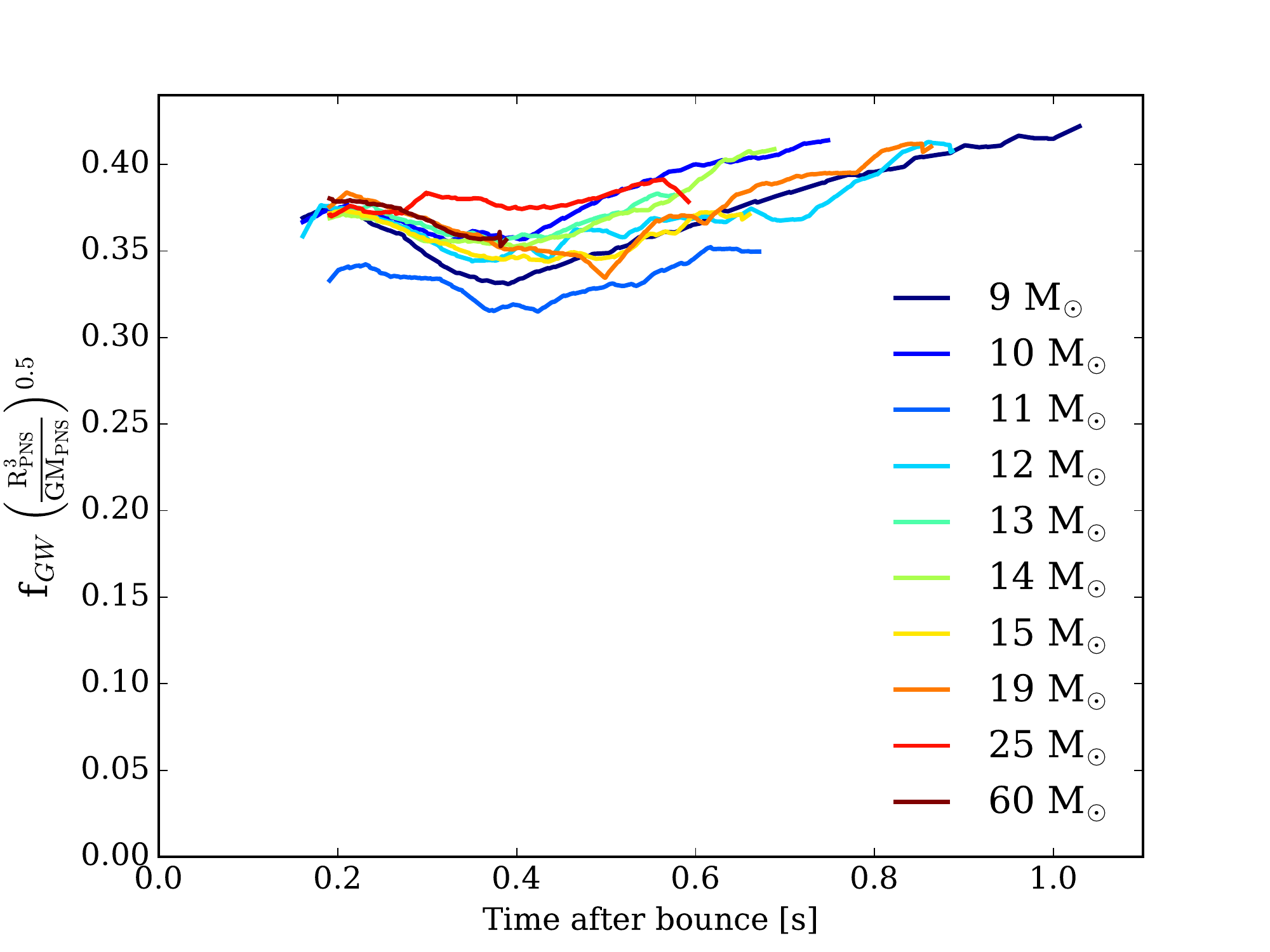}
    \caption{The gravitational wave fundamental mode frequency normalized by PNS dynamical time, a function of the PNS radius and mass, for the various progenitors plotted versus time after bounce (in seconds). We see little variation, approximately $\sim$10\% in the normalized fundamental frequency, which is itself approximately 0.35 $\times$ the dynamical time 500 ms after bounce and tends to increase with time. See text for a discussion.}
    \label{fig:freq_rad_mass}
\end{figure*}

The $f$-mode gravitational waves depend approximately upon the average density of the PNS (\citealt{muller2013,sotani2019,torres2019}). In Fig.\,\ref{fig:freq_rad_mass}, we plot the gravitational wave fundamental mode frequency normalized by the PNS dynamical time, depending on the PNS mass and radius alone.  The result is largely progenitor-independent (roughly 0.35 divided by the dynamical time), with only $\sim$10\% variation by progenitor mass, suggesting that observations of gravitational wave frequencies from a galactic event will constrain the PNS mass and radius. Independent measurement of the PNS gravitational mass will further constrain the PNS radius, providing insight into the PNS mass-radius relation and nuclear equation of state. 

\section{Conclusions}\label{sec:conc}

We have provided in this paper the time series and angular distributions of the neutrino and gravitational-wave 
emissions of eleven state-of-the-art 3D non-rotating models of core-collapse supernovae and explored possible correlations 
between these signatures and the real-time dynamics of the shock and the proto-neutron-star core.\footnote{The neutrino and gravitational wava data are available from the authors upon request.} This is the largest 
set of high-fidelity 3D simulations yet performed and reveals the global characteristics and general systematics
for a wide range of available progenitor structures.  Identifications of the predicted temporal fluctuations in 
these emissions in detectors on Earth can be used to constrain core and explosion dynamics in real time before, 
during, and after the supernova explosion is underway. We find that the neutrino emissions of non-rotating models
retain a good degree of isotropy on average, but with instantaneous excursions about the mean inferred luminosity 
in a given direction of as much as $\pm$20\%.  Dipolar angular RMS variations in the neutrino emissions are 
generally restricted to $\sim$10\% of the mean after the non-linear turbulent phase is reached and after 
the supernova is launched (if it is).  Most explosions involve in their first phases simultaneous explosion 
and accretion, with a wasp-waist structure and unsymmetric-dipolar explosive ejection \citep{burrows_2019}. 
The deviation from isotropy is least for the ``$\nu_{\mu}$"-type neutrinos. The temporal variations in 
the neutrino emissions reflect the motions of the standing accretion shock before explosion, which itself 
correlates with the temporal and angular variations in the mass accretion rate through the shock. 
In particular, we identify a temporal correlation between the dipole component of the shock surface
and the corresponding quantities for the accretion rate and the neutrino emissions.  We also find that the vector 
direction of the dipole of the Y$_\mathrm{e}$ distribution in the inner region of proto-neutron-star convection 
and of that of the LESA are highly correlated, particularly after explosion.  Furthermore, we witness the LESA phenomenon
in all our models, though with distinctive differences in magnitude and temporal evolution. 

We find, not unexpectedly, that the time series of the neutrino signatures at a detector 
bear the stamp of the hydrodynamics of shock motion before explosion and of episodic mass 
accretion after explosion. Determination of the characteristic timescales of these temporal fluctuations 
in the neutrino signals can be used to determine the timescales and behavior of shock motion. In particular,
if there is a SASI (which we see for only non-exploding models, or over a short time span for models with late explosions),
it is reflected in a correponding temporal variation at the SASI frequency in the neutrino signal and at twice the 
SASI frequency in the gravitational-wave signal. For our non-rotating model set, though the instantaneous 
values of the different gravitational strain polarizations, h$_{+,\times}$, can be out of phase, on average they 
are similar in magnitude and frequency content.  However, their instantaneous angular variation, both for a given polarization and between
polarizations, can be quite large.  The relative magnitude of the different polarizations in the 
initial phase of gravitational radiation, associated with the distinct prompt-convection
phase, bears the stamp of the angular character of the initial perturbations.  Angular emission patterns and 
differences between polarizations would be even more pronounced for rapidly-rotating models for which a given 
preferred direction is set and for which polarization differences in the angular emission pattern won't
average out \citep{hayama2018}.  

For our entire set of 3D models, we find strong connections between the cumulative neutrino energy losses, the radius
of the proto-neutron star, and the $f$-mode frequency of the gravitational wave emissions.  All these quantities are
monotonic with time after bounce and measurement of, for instance, the $f$-mode frequency or the cumulative 
neutrino energy loss can be used to constrain the others. When physically normalized, the progenitor-to-progenitor 
variation in any of these reduced quantities is no more than $\sim$10\%.  Moreover, the $f$-mode frequency, times the instantaneous physical dynamical time, is independent of time after bounce to better than $\sim$10\%. This implies that 
simultaneous measurement of gravitational waves and neutrinos from a given supernova event can be used together 
to extract real physical quantities of the core from which the supernova explosion is launched.  Hence, and 
importantly, the neutrino data can aid in the interpretation of the gravitational-wave data, and vice versa. 
Since both the distance and spectral type of a galactic core-collapse progenitor are likely to be determined
quickly, constraining possible core structures and enabling the determination of absolute neutrino and gravitational-wave
powers, the additional correlations we highlight here will enhance the potential scientific return from a galactic event.

In this paper, we have mined our recent extensive suite of 3D supernova simulations to explore some of
the correlations between core dynamics and its temporal evolution and the dominant neutrino and 
gravitational-wave signatures upon which this dynamics is stamped.  The next step will be to filter our
emission predictions through detector pipelines, including in the case of neutrinos the effect of neutrino 
oscillations \citep{seadrow2018}. Whether the features we have predicted can be discerned at Earth
will depend upon supernova distance and detector characteristics. However, whatever detector configurations exist and are online when the next galactic core-collapse supernova explodes, we hope we have demonstrated 
with this paper that modern 3D simulations, incorporating the necessary physical realism, can be used profitably 
and in detail to inform the interpretation of such a marvelous astronomical opportunity, when next it arises.


\section{Acknowledgments}
The authors are grateful for useful discussions with Sean Couch and Hiroki Nagakura and insights from Josh Dolence and Aaron Skinner. We acknowledge support from the U.S. Department of Energy Office of Science and the Office of Advanced Scientific Computing Research via the Scientific Discovery through Advanced Computing (SciDAC4) program and Grant DE-SC0018297 (subaward 00009650) and support from the U.S. NSF under Grants AST-1714267 and PHY-1144374.  The authors also acknowledge generous HPC allocations from the DOE/ASCR INCITE program under Contract DE-AC02-06CH11357; Blue-Waters PRACs under awards OCI-0725070, ACI-1238993, TG-AST170045, and OAC-1809073; and an XSEDE/Stampede2 award (under ACI-1548562). Furthermore, the authors employed computational resources provided by the TIGRESS high performance computer center at Princeton University, which is jointly supported by the Princeton Institute for Computational Science and Engineering (PICSciE) and the Princeton University Office of Information Technology, and acknowledge our continuing allocation at the National Energy Research Scientific Computing Center (NERSC), which is supported by the Office of Science of the US Department of Energy (DOE) under contract DE-AC03-76SF00098. DR cites partial support as a Frank and Peggy Taplin Fellow at the Institute for Advanced Study.

\clearpage

\clearpage
\bibliographystyle{mnras}
\bibliography{References}

\end{document}